\documentclass[aps,pra,reprint,superscriptaddress,longbibliography]{revtex4-2}
\usepackage[utf8]{inputenc} 
\usepackage[top=2cm, bottom=2cm, left=1.5cm, right=1.5cm]{geometry} 
\usepackage{adjustbox}
\usepackage{gensymb} 
\usepackage{amsmath}
\usepackage{upgreek} 
\usepackage{graphics} 
\usepackage{graphicx} 
\usepackage[colorlinks,bookmarks=false,linkcolor=black,urlcolor=blue, citecolor=black]{hyperref} 
\usepackage[english]{babel} 
\usepackage{wrapfig} 
\usepackage[T1]{fontenc} 
\usepackage{array} 
\usepackage{cleveref}
\usepackage{booktabs}
\usepackage{textcomp}
\usepackage{physics}
\usepackage{appendix}
\usepackage{verbatim}

\newcommand{\notleftright}{\mathrel{\ooalign{$\leftrightarrow$\cr\hidewidth$/$\hidewidth}}}

\begin{document}

\title{Characterization of the electronic ground state of He$_2^+$ by high-resolution photoelectron spectroscopy}

\author{M. Holdener}
    \affiliation{\small Department of Chemistry and Applied Biosciences, ETH Zürich, 8093 Zürich, Switzerland}
\author{V. Wirth}
    \affiliation{\small Department of Chemistry and Applied Biosciences, ETH Zürich, 8093 Zürich, Switzerland}
\author{N. A. Shahin}
    \affiliation{\small Department of Chemistry and Applied Biosciences, ETH Zürich, 8093 Zürich, Switzerland}
\author{M. Beyer}
    \affiliation{\small Department of Physics and Astronomy, Vrije Universiteit Amsterdam, The Netherlands}
\author{F. Merkt}
    \affiliation{\small Department of Chemistry and Applied Biosciences, ETH Zürich, 8093 Zürich, Switzerland}
    \affiliation{\small Department of Physics, ETH Zürich, 8093 Zürich, Switzerland}
    \affiliation{\small Quantum Center, ETH Zürich, 8093 Zürich, Switzerland}

\date{\small\today}

\begin{abstract}
     Excluding the very shallow potential minimum of the electronic ground state, all bound electronic states of He$_2$ have Rydberg character. 
     Their potential-energy functions are very similar to the potential-energy functions of the He$_2^+$ states to which the Rydberg series converge. 
     Photoionization and electron-impact ionization of the metastable $a~^3\Sigma_u^+$ state of He$_2$ are thus characterized by diagonal Franck-Condon factors and only provide access to low vibrational levels of the He$_2^+$ $X^+$ $^2\Sigma_u^+$ electronic ground state. 
     For this reason, hardly any experimental information is available on the excited vibrational levels of He$_2^+$. We report on a measurement, by high-resolution photoelectron spectroscopy, of 17 vibrational levels of the $X^+$ $^2\Sigma_u^+$ state of $^4$He$_2^+$, with vibrational quantum number $v^+$ ranging from 3 to 19 and covering more than 95\% of the potential well. 
     To access these states, we exploit a hump in the potential-energy function of the $c~^3\Sigma_g^+$ state, which has vibrational wavefunctions extending to large internuclear distance by quantum-mechanical tunneling through the potential barrier. 
     Combining these new results with data available on the lowest ($v^+=0-2$) and highest ($v^+=22$, 23) vibrational levels of $^4$He$_2^+$, we derive a full map of the rovibrational level structure of He$_2^+$ and use it to determine, in a least-squares fit, an empirical effective potential-energy function that describes all experimental data within their uncertainties. This potential-energy function is used to calculate the positions of the 409 bound rovibrational levels and the positions and widths of the 74 shape resonances of the $X^+$ $^2\Sigma_u^+$ electronic ground state of $^4$He$_2^+$. The dissociation energy of He$_2^+$ is determined to be $D_{\rm e}=19\,956.10(10)$~cm$^{-1}$ [$D_0(^4{\rm He}_2^+)=19\,101.29(10)$~cm$^{-1}$].
\end{abstract}
\maketitle
\newpage
{\small Accepted for publication in Physical Review A, 2025 Jul. 25, DOI: \url{https://doi.org/10.1103/w95m-w7hc}}

\section{Introduction}

    The He$_2^+$ molecular cation is one of very few molecular systems for which highly accurate first-principles calculations are possible. 
    With only three electrons, He$_2^+$ has been an important reference quantum-mechanical system since the early days of modern quantum mechanics \cite{majorana31a,pauling33a}. 
    First experimental information on the energy-level structure of He$_2^+$ were derived from spectra recorded in helium tubes \cite{curtis13a,goldstein13a}, which contain spectral structures that were soon interpreted as corresponding to Rydberg series of He$_2$ \cite{fowler15a,mulliken26a,weizel29a,weizel29b,weizel29c}. 
    Extrapolations of these Rydberg series with the Rydberg formula indicated that He$_2^+$ is a strongly bound molecule and led to the first determinations of the rovibrational level structure of the ground state of He$_2^+$.  
    These observations were confirmed theoretically by Majorana \cite{majorana31a} and Pauling \cite{pauling33a}, who applied the valence-bond theory of Heitler and London to calculate the dissociation energy (1.41~eV \cite{majorana31a}; 2.47~eV \cite{pauling33a}) and equilibrium distance (1.16~\AA \ \cite{majorana31a}; 1.085~\AA \ \cite{pauling33a}) of the $X^+$ $^2\Sigma_u^+$ ground state of He$_2^+$ ($X^+$ state hereafter).

    100 years after these pioneering investigations, the theoretical treatment of the level structure of He$_2^+$ has been considerably extended. 
    The latest theoretical investigations report electronic-structure calculations carried out with very large basis sets \cite{reagan63a,liu71a,ackermann91a,cencek95a,tung12a,matyus18a,gebala23a,bauschlicher89a, kedziera22a,matyus25a, bawagan97a} and provide equilibrium constants, dissociation energies, and in some cases even full sets of rovibrational energies.  
    Recently, even nonadiabatic, relativistic and quantum-electrodynamics contributions to level energies have been evaluated \cite{tung12a,matyus18a,ferenc20a,matyus25a}, providing extensive reference material for comparison with experimental data.
    
    $^4$He$_2^+$ and $^3$He$_2^+$ are homonuclear diatomic molecules and do not have electric-dipole-allowed pure-rotational and rovibrational transitions. 
    The only transitions observed in these molecular ions are microwave electronic transitions between the highest bound rovibrational levels $(v^+,N^+)= (22,5), (23,1)$ and (23,3) of the $X^+$ electronic ground state and the weakly bound (0,0), (0,2), (0,4), (1,0) and (1,2) levels of the $A^+$ $^2\Sigma_g^+$ first excited electronic state~\cite{carrington95b}. 
    Nine $R$-Type vibrational transitions of $^4$He$^3$He$^+$, which has a permanent electric dipole moment, have been observed in the IR, connecting $N^+=1,3-11$ rotational levels of the vibrational ground state ($v^+=0$) with $N^+=2,4-12$ rotational levels of the first excited vibrational level ($v^+=1$) \cite{yu87a,yu88b}. 
    
    Most of the spectroscopic data available on He$_2^+$ have been derived  from studies of the Rydberg spectrum of He$_2$ from the metastable $(1{\rm s}\sigma_g)^2(1{\rm s}\sigma_u)^1(2s\sigma_g)^1$ $a~^3\Sigma_u^+$ state of He$_2$ ($a$ state hereafter) in combination with Rydberg-series extrapolation using multichannel quantum defect theory \cite{ginter84a,sprecher14a,semeria16a,jansen16a,jansen18a,jansen18b,semeria20a} or from rotationally resolved photoelectron spectra \cite{raunhardt08a,motsch14a}. 
    These data consist of the term values of the $v^+=0-2,N^+=1-23$ rovibrational levels of $^4$He$_2^+$ and of the $v^+=0,N^+=0-11$ rovibrational levels of $^3$He$_2^+$. 
    In addition, several long-lived quasibound levels of $^4$He$_2^+$, $^4$He$^3$He$^+$ and $^3$He$_2^+$ have been characterized in measurements of momentum distributions of He$^+$ fragments \cite{maas76a}.
    
    The main experimental knowledge gap in the energy-level structure of He$_2^+$ concerns the vibrational levels in the range of quantum number $v^+$ between 3 and 21. 
    Such levels are difficult to access experimentally. Firstly, they are not significantly populated in the ion sources used in high-resolution spectroscopic studies; secondly, they are not easily accessible from the low vibrational levels ($v^\prime \leq 2)$ of the metastable $a$ state of He$_2$ that are populated in electric discharges. 
    Indeed, the $a$ state, having itself Rydberg character, has a potential-energy function that is almost identical to that of the He$_2^+$ ion. Consequently, the Franck-Condon factors for ionization or for the excitation to higher-lying  Rydberg states of He$_2$ favor transitions with $\Delta v=0$.
    Finally, the electronic ground state of He$_2$ is essentially repulsive, with only a single extremely weakly bound rovibrational level \cite{grisenti00a}, and all other electronic states of He$_2$ have Rydberg character \cite{herzberg87a}.  
    	
    \begin{figure}[!t]
        \begin{center}
            \includegraphics[width=\columnwidth]{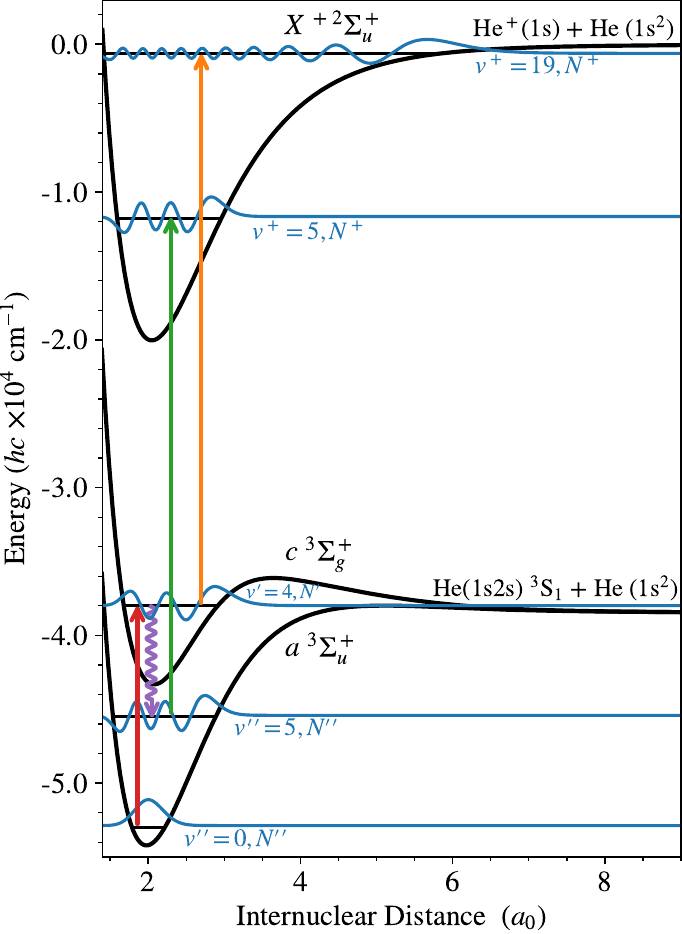}
            \caption{\label{fig:FIG1_PEC_schemes}
            Potential energy functions of the two triplet states associated with the He(1s2s) $^3$S$_1$ + He(1s$^2$) $^1$S$_0$ dissociation limit of He$_2$ (from Ref.$ $ \cite{yarkony89a}) and of the electronic ground state of He$_2^+$ (from Ref.$ $ \cite{tung12a}). 
            Black horizontal lines and blue functions represent the eigenvalues and eigenfunctions, respectively, of the labeled vibronic states. 
            The red arrow indicates the transition to the $c~^3\Sigma_g^+$ ($v^\prime = 4$) state from the initial $a~^3\Sigma_u^+$ ($v^{\prime\prime} = 0$) state, which can be used to populate vibrationally excited states of the $a~^3\Sigma_u^+$ state by radiative decay (purple arrow).
            The green and orange arrows represent photoionizing transitions used to access vibrationally excited levels of the $X^+$ $^2\Sigma_u^+$ ($v^+$) states of He$_2^+$ from the $a~^3\Sigma_u^+$ ($v^{\prime\prime} = 5$) and $c~^3\Sigma_g^+$ ($v^\prime = 4$) states, respectively.}
        \end{center}
    \end{figure}
    
    We report here on the experimental characterization, by high-resolution photoelectron spectroscopy, of numerous rovibrational levels of the $X^+$ electronic ground state of $^4$He$_2^+$ with vibrational quantum number $v^+$ up to 19. 
    To access these states, we exploit the interaction  between the potential functions of the lowest member of the $n$p $^3\Sigma_g^+$ Rydberg series converging on the strongly bound He$_2^+$ $X^+$ ionic ground state and the lowest member of the $n$s $^3\Sigma_g^+$ Rydberg series converging on the essentially repulsive first excited He$_2^+$ $A^+$ $^2\Sigma_g^+$ state. 
    This interaction leads to an avoided crossing \cite{guberman72a} and to a potential hump in the potential-energy curve of the $c~^3\Sigma_g^+$ state of He$_2$ ($c$ state hereafter) and to the predissociation of its $v=3$ and 4 vibrational levels by quantum-mechanical tunneling. 
    This effect, which was analyzed in detail by Bjerre and coworkers \cite{lorents89a,kristensen90a}, provides access to a much larger range of internuclear distances and is exploited in the present work to study high vibrational levels of the $X^+$ state of He$_2^+$.
    
    The excitation schemes we use for this purpose and the relevant potential-energy functions are illustrated in Fig.~\ref{fig:FIG1_PEC_schemes}. 
    Starting from the low-lying $v^{\prime\prime}=0$ and 1 vibrational levels of the $a$ state of He$_2$, which we produce in an electric discharge, we access the $v^\prime=3$ and 4 levels of the $c$ state. 
    These levels are used as intermediate levels to increase the internuclear separation in subsequent photoexcitation and photoemission steps for the study of high vibrational levels of the $X^+$ state. 
    The data on the $v^+=3-1$9 vibrational levels of the $X^+$ electronic ground state of He$_2^+$ obtained in this way are combined with previous data on the $v^+=0-2$ and $v^+=22$, 23 levels to derive, in a least-squares fit, an effective potential-energy function for the $X^+$ state of He$_2^+$. 
    This function describes the available experimental data within their uncertainties and is used to predict the positions and widths of the shape resonances of $^4$He$_2^+$.

\section{Experimental setup and methods}\label{sec:experimental}

    \begin{figure}[!t]
        \begin{center}
            \includegraphics[width=\columnwidth]{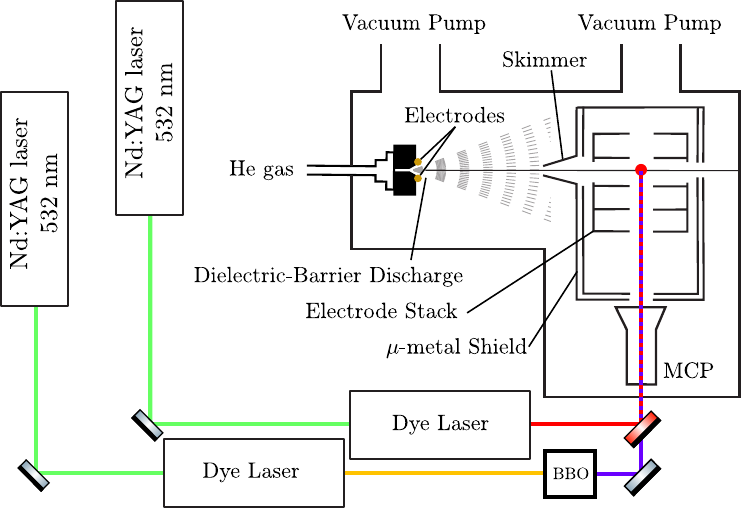}
            \caption{\label{fig:FIG2_experimental_setup}
            Schematic representation (not to scale) of the experimental setup used to record high-resolution photoelectron and photoionization spectra of He$_2$ from the metastable $a$ state and consisting of a laser system, a pulsed valve equipped with a dielectric-barrier discharge, a magnetically shielded photoexcitation region and a linear ion/electron time-of-flight spectrometer. 
            Nd:YAG = neodymium-doped yttrium-aluminum-garnet; BBO = beta-barium-borate crystal; MCP = microchannel-plate detector.}
        \end{center}
    \end{figure}

    To access the high vibrational levels ($v^+>2$) of the $X^+$ state of He$_2^+$, two different excitation schemes, depicted in Fig.~\ref{fig:FIG1_PEC_schemes}, were employed. 
    Both schemes rely on the initial excitation of He$_2$ from the $a(v^{\prime\prime} = 0)$ state to the $c(v^\prime = 3,4)$ states. 
    In the first scheme, the excited He$_2$ molecules are allowed to radiatively decay to the $a(v^{\prime\prime} = 3-5)$ states, which are then used as starting levels to record pulsed-field-ionization zero-kinetic-energy photoelectron (PFI-ZEKE PE) spectra \cite{muellerdethlefs98a} of excited vibrational levels of the $X^+$ state of He$_2^+$ with $v^+=3-5$. 
    In the second scheme, selected rotational levels of the $c(v^\prime = 3,4)$ states are used as intermediate levels to record PFI-ZEKE PE spectra of very high vibrational levels of the $X^+$ state of He$_2^+$ with $v^+$ up to $19$.

    The experimental setup is represented schematically in \Cref{fig:FIG2_experimental_setup}. 
    Two commercial dye lasers, each pumped by the second harmonic of a Q-switched neodymium-doped yttrium-aluminum-garnet (Nd:YAG) pump laser at a repetition rate of 25~Hz, were employed to generate the frequencies required for the excitation schemes. 
    The lasers produced radiation with a bandwidth of 0.1 cm$^{-1}$. 
    The first laser, used to drive the $c(v^\prime = 3\ {\rm and}\ 4, N^\prime)\leftarrow a(v^{\prime\prime} = 0\ {\rm and}\ 1, N^{\prime\prime})$ transitions of He$_2$, was operated in the visible range at wavelengths between 600 to 680 nm with pulse energies in the range from 0.2 to 0.4 mJ. 
    To access the region of the $X^+(v^+=3-19, N^+)$ ionization thresholds, the second laser was scanned in the UV range from 270 to 375~nm with pulse energies ranging from 0.2 to 1.5~mJ after doubling the frequency of the dye-laser output (fundamental wavelength in the range $540-730$~nm) using a beta-barium-borate (BBO) crystal. 
    Both laser beams were combined using a dichroic mirror and transmitted into the vacuum chamber through a fused-silica window. 
    The frequencies of both dye lasers were calibrated using commercial wavemeters (HighFinesse WS6-600 and a WS7-60) with absolute accuracies of 600 and 60 MHz, respectively.

    The laser beams intersected a supersonic beam containing He$_2$ in its $a$ metastable state at right angles. 
    These molecules were produced by a dielectric-barrier discharge located near the orifice of a pulsed valve. The valve, operated at a stagnation pressure of 12 bar of pure He gas, was cooled to 77 K to generate a supersonic expansion with a forward velocity of $\sim$1000 m/s. 
    The central part of the molecular beam was selected by a 3-mm-diameter skimmer positioned 40 cm downstream of the valve orifice, which reduced the transverse velocity and the Doppler width of the $c\leftarrow a$ transitions. 
    The vibrational and rotational degrees of freedom of the metastable $a$ state were not efficiently cooled in the supersonic expansion, as discussed further in Section \ref{subsec:results_c_state}.

    Photoexcitation of He$_2$ to the region of the He$_2^+(v^+,N^+)$ ionization thresholds took place on the axis of a 5.8-cm-long stack of five resistively coupled cylindrical electrodes designed for the application of pulsed homogeneous electric fields. 
    This electrode stack was surrounded by two concentric $\mu$-metal shields to suppress stray magnetic fields. Depending on the polarity of the applied electric fields, either the He$_2^+$ ions or the electrons produced by photoionization were extracted along a flight tube for detection at a microchannel-plate (MCP) detector. 
    
    To record spectra of the $c(4, N^\prime)\leftarrow a( 0, N^{\prime\prime})$ transitions of He$_2$, $(1 + 1^\prime)$ resonance-enhanced two-photon ionization spectroscopy was employed. The wavenumber of the second (UV) laser was kept fixed at a position well above the first ionization threshold of the metastable $a$ state and the spectra were obtained by monitoring the He$_2^+$ ion signal as a function of the wavenumber of the first (visible) laser following extraction to the MCP detector using an electric-field pulse of 120~V/cm.

\begin{figure*}[!t]
        \begin{center}
            \includegraphics[width=\linewidth]{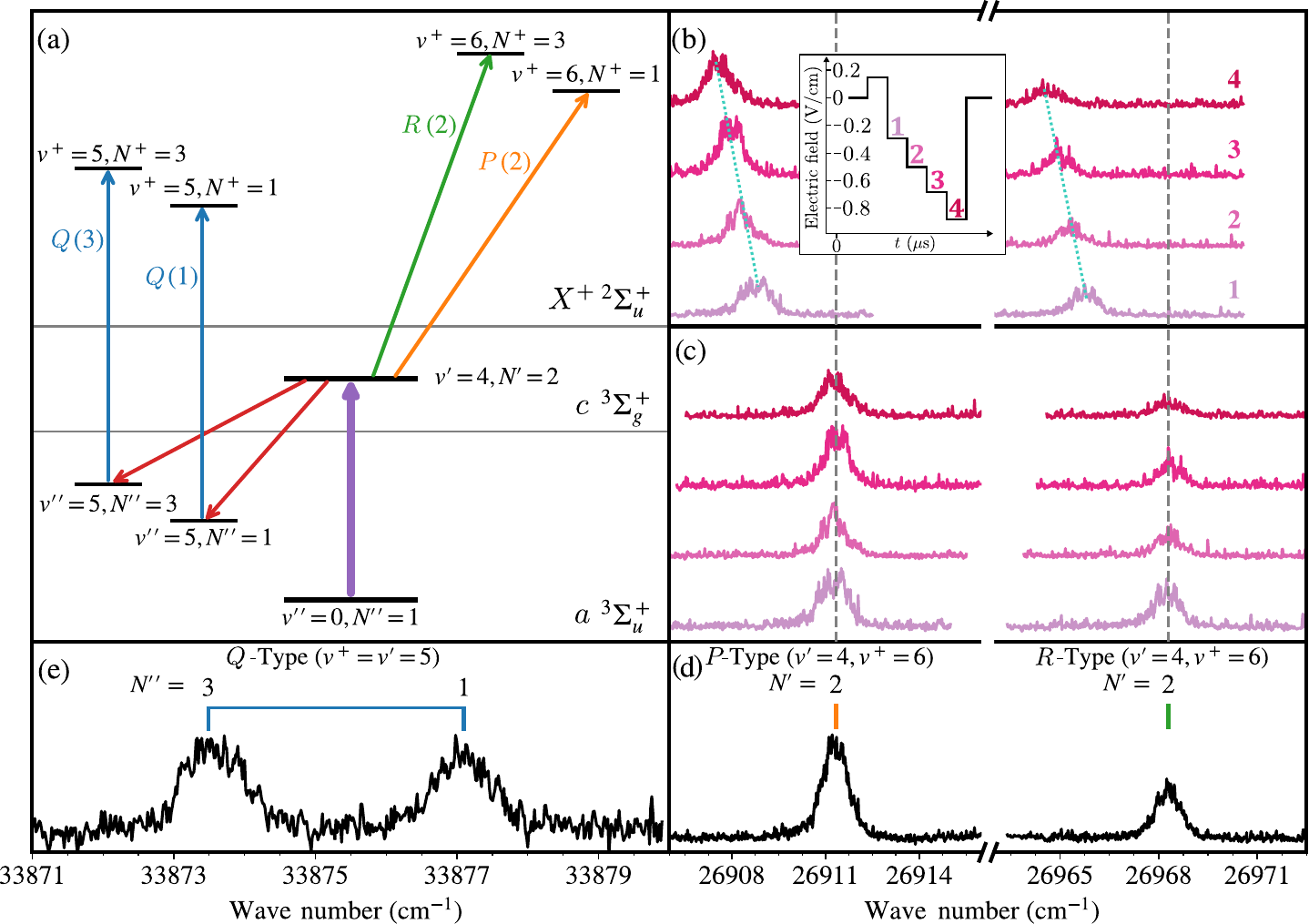}
            \caption{\label{fig:FIG3_multiPanel_scheme_ZekeShift_Modified} 
            (a) Energy-level diagram of He$_2$ illustrating the two schemes used to record PFI-ZEKE PE spectra of vibrationally excited states of He$_2^+$. 
            (b), (c) PFI-ZEKE PE spectra of the transitions from the $c~^3\Sigma_g^+(v^\prime = 4, N^\prime = 2$) state of He$_2$ to the $X^+$ $^2\Sigma_u^+$ ($v^+ = 6, N^+ = 1,3$) states of He$_2^+$ (orange and green arrows in (a)) recorded before (b) and after (c) correcting for the field-induced shifts of the ionization thresholds induced by the sequence of four electric-field pulses displayed in the inset. 
            (d) Weighted sum of the spectra displayed in (c). (e) Field-corrected PFI-ZEKE PE spectrum of the transitions from the $a~^3\Sigma_u^+$ ($v^{\prime\prime} = 5, N^{\prime\prime} = 1,3$) state of He$_2$ to the $X^+$ $^2\Sigma_u^+$ ($v^+ = 5, N^+ = 1,3$) state of He$_2^+$ (blue arrows in (a)). See text for details.
            }
        \end{center}
    \end{figure*}

    PFI-ZEKE PE spectroscopy \cite{muellerdethlefs98a} was used to study the rovibrational levels of the $X^+$ state of He$_2^+$. The spectra were recorded as a function of the wavenumber of the second (UV) laser by monitoring the electron signal resulting from the delayed pulsed field ionization of long-lived high Rydberg states of He$_2$ with principal quantum number $n\geq 100$ located just below the different $X^+(v^+,N^+)$ ionization thresholds.
    These long-lived Rydberg states were field-ionized using sequences of weak electric-field pulses, and the resulting electrons were extracted toward the MCP detector. 
    A sequence of five electric-field pulses $[(0.17, -0.35, -0.52,-0.69, -0.86)$ V/cm], displayed in the inset of Fig.~\ref{fig:FIG3_multiPanel_scheme_ZekeShift_Modified}b, was used to record the PFI-ZEKE PE spectra displayed in panels b-e of this figure, achieving a spectral resolution of 0.4 cm$^{-1}$ (full width at half maximum).
    The first pulse, of positive polarity, served the purpose of sweeping prompt electrons out of the photoexcitation region. The remaining four pulses (labeled 1-4, in Fig.~\ref{fig:FIG3_multiPanel_scheme_ZekeShift_Modified}b) were used to generate four distinct pulsed-field-ionization signals. These signals were monitored separately to obtain four PFI-ZEKE PE spectra for each laser scan. 

    Figure~\ref{fig:FIG3_multiPanel_scheme_ZekeShift_Modified} illustrates the procedure followed to obtain PFI-ZEKE PE spectra with this five-pulse sequence with the examples of the transitions to the $v^+=5$ and $6$ vibrational levels of the $X^+$ state of He$_2^+$ from the $a(5, N^{\prime\prime} = 1,3$) and $c(4,N^\prime=2)$ states of He$_2$, respectively.
    The relevant levels of He$_2$ and He$_2^+$ are depicted in Fig.~\ref{fig:FIG3_multiPanel_scheme_ZekeShift_Modified}a, where the photoionizing transitions are color-coded in blue for the $X^+\leftarrow a$ transition and in green and orange for the $X^+\leftarrow c$ transition.
    The PFI-ZEKE PE spectra of the $X^+(6, N^+) \leftarrow c(4,2)$ thresholds are depicted in Fig.~\ref{fig:FIG3_multiPanel_scheme_ZekeShift_Modified}b, where the corresponding pulsed-field labels are indicated on the right.
    The successive pulses of the sequence induce the field ionization of Rydberg states of increasing binding energy so that the corresponding PFI-ZEKE PE spectra gradually shift to lower wavenumbers. 
    The field-induced shifts can be calculated and corrected for by following the procedure introduced in Ref.$ $ \cite{hollenstein01a}. After correction, the spectra recorded with the different pulses of the sequence are all perfectly aligned and the line positions correspond to the field-free ionization thresholds. 
    A weighted sum of the field-corrected spectra leads to the spectrum displayed in Fig.~\ref{fig:FIG3_multiPanel_scheme_ZekeShift_Modified}d, which combines the advantages of a high spectral resolution and an improved signal-to-noise ratio, as first demonstrated in Ref.~\cite{mollet13a}. 
    
    Applying the same electric-field pulse sequence when recording the PFI-ZEKE PE spectrum of the $X^+(5, N^+ = 1,3)\leftarrow a(5, N^{\prime\prime} = 1,3)$ transitions led to the spectrum presented in Fig.~\ref{fig:FIG3_multiPanel_scheme_ZekeShift_Modified}e. 
    The increase in signal-to-noise ratio resulting from this procedure turned out to be crucial to measure spectra of the very weak transitions to high-lying vibrational levels of the $X^+$ state of He$_2^+$.
    To enhance the accuracy of the field-correction procedure, the same pulse sequences were also used to record the strong $X^+ (v^+ = 0, 1) \leftarrow a(v^{\prime\prime} = 0, 1)$ transitions. The field-induced shifts caused by the different pulses were calibrated by making reference to the very accurate $X^+(v^+ = 0,1)$ ionization thresholds determined by Semeria {\it et al.} \cite{semeria20a}. 
    We estimate the absolute accuracy of the thresholds determined in the present study to be $0.1$ cm$^{-1}$. 

\section{Results and discussion}

    \subsection{The $c~^3\Sigma_g^+$ $(v^{\prime} = 3,4)\leftarrow a~^3\Sigma_u^+$ $(v^{\prime\prime} = 0,1)$ band system and tunneling in the intermediate $c~^3\Sigma_g^+$ $(v^{\prime} = 3,4)$ states}\label{subsec:results_c_state}

\begin{figure*}[ht]
            \begin{center}
                \includegraphics[width=\linewidth]{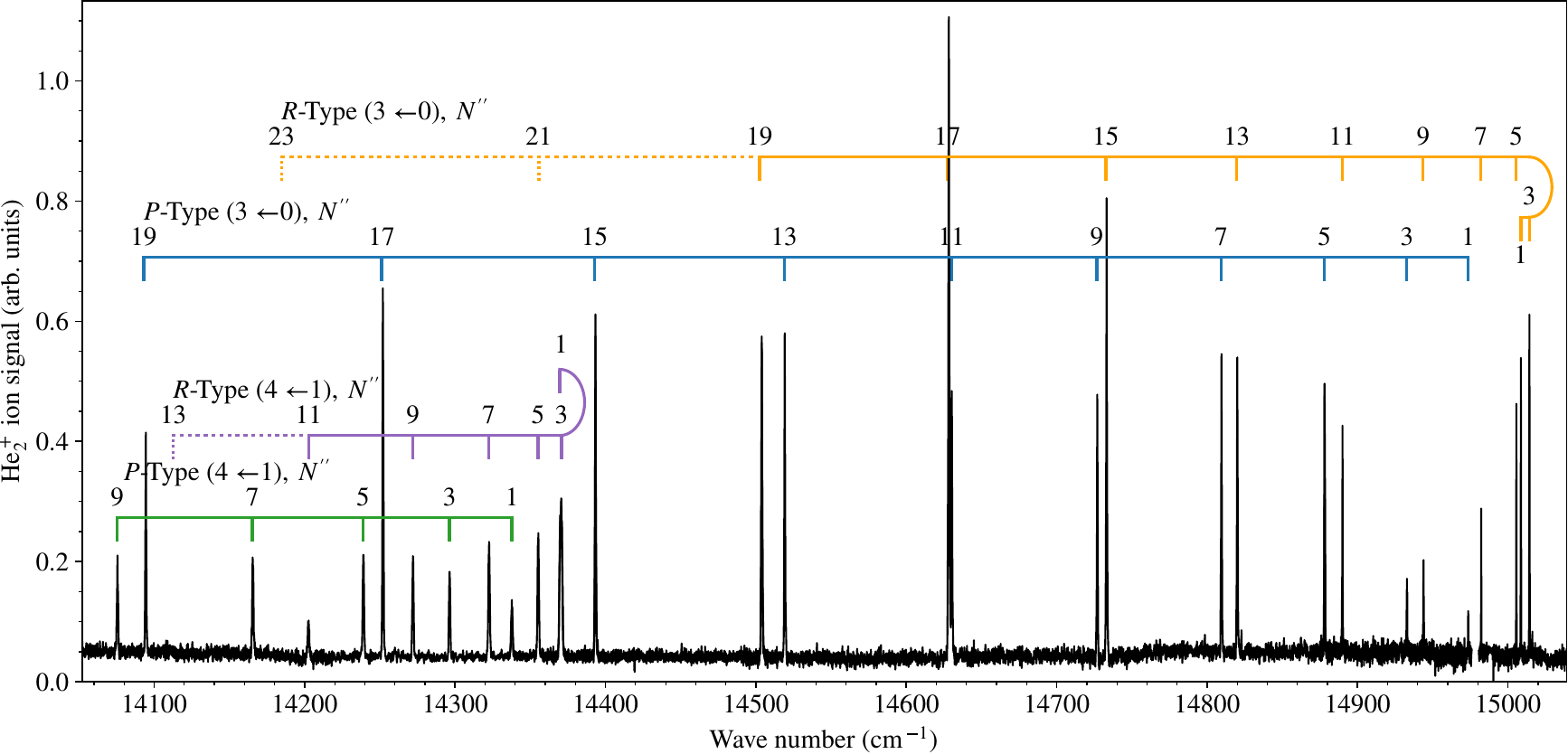}
                \caption{\label{fig:FIG4_Rempi_a_c} 
                ($1+1^\prime$) resonance-enhanced two-photon-ionization spectrum of the overlapping $(4-1)$ and $(3-0)$ bands of the $c~^3\Sigma_g^+ \leftarrow a~^3\Sigma_u^+$ transition of He$_2$ obtained by merging several segments recorded separately and under slightly different experimental conditions.
                Vertical full (dashed) tick marks along the full (dashed) horizontal assignment bars indicate the positions of observed (predicted but not observed) transitions.
                }
            \end{center}
        \end{figure*}

        Individual rotational levels of the $c(v' = 3, 4)$ states of He$_2$ served as intermediate levels to access high vibrational levels of the $X^+$ state of He$_2^+$ from low vibrational levels of the $a$ state, either by resonance-enhanced ($1+1^\prime$) two-photon ionization or through an absorption-emission-absorption sequence, as explained in Section~\ref{sec:experimental}. 
        The ($1+1^\prime$) resonance-enhanced two-photon ionization spectrum of He$_2$ in the region of the overlapping $(4- 1)$ and $(3- 0)$ bands of the $c\leftarrow a$ transitions displayed in Fig.~\ref{fig:FIG4_Rempi_a_c} gives an overview of the $c$ state rotational levels that were accessible in the present study. This spectrum was recorded in several segments and was not corrected for the varying laser intensity. Consequently, the intensity distribution only provides qualitative information on the relative populations of the $a$ state rovibrational levels.  
        
        The $a$ and $c$ states of He$_2$ are both triplet ($S = 1$) states and each rotational level, with rotational quantum number $N$, splits into three components of total angular momentum $\vec{J} = \vec{S} + \vec{N}$ and $J = N, N \pm 1$. 
        Electric-dipole transitions between the $a(v^{\prime\prime}, N^{\prime\prime}, J^{\prime\prime})$ and $c(v^\prime, N^\prime, J^\prime)$ levels obey the selection rule $\Delta J = J^\prime - J^{\prime\prime} = 0, \pm 1, $ $0\notleftright 0 $.
        Our experimental resolution was not sufficient to observe the fine-structure splittings in the $c-a$ band system resulting from  the spin-spin and spin-rotation interactions. The rotational structure was treated (and labeled) in Hund's angular-momentum coupling case (b) using the rotational angular momentum number $N$ (i.e., $N^{\prime\prime}$ and $N^{\prime}$ for the $a$ and $c$ states, respectively).
        Because the $^4$He$^{2+}$ nuclei are bosons with zero nuclear spin ($I = 0$), the Pauli principle requires the total wavefunction to be symmetric under exchange of the two alpha particles. 
        Consequently, only levels with even (odd) $N$ values are populated in states of $\Sigma_g^+$ ($\Sigma_u^+$) symmetry and the allowed electric-dipole rovibronic transitions of the $c-a$  band system must obey the selection rule $\Delta N = N^\prime - N^{\prime\prime} = \pm 1, \pm 3$.
        The transitions with $\Delta N = \pm 3$ are much weaker than those with $\Delta N=\pm 1$ and could not be detected. 
        
        The $c-a$ $(3-0)$ and $(4-1)$ bands depicted in Fig.~\ref{fig:FIG4_Rempi_a_c} both consist of two rotational branches, an $R$-Type branch corresponding to $\Delta N = N^\prime - N^{\prime\prime} = +1$ (with orange and violet assignment bars for the $(3-0)$ and $(4-1)$ bands, respectively) and a $P$-Type branch ($\Delta N = -1$, with blue and green assignment bars for the $(3-0)$ and $(4-1)$ bands, respectively).
        The assignment bars drawn as full lines designate the observed transitions whereas the dashed bars indicate the predicted positions of the lines that could not be detected. Particularly striking features of these rotational branches are their abrupt ends, at $N^{\prime}=20$ and 12 for the $(3-0)$ and $(4-1)$ bands, respectively. 
        
        Spectra of other bands of He$_2$ recorded in the present study (not shown) as well as in previous studies of He$_2$ \cite{raunhardt08a,motsch14a,jansen18b} using similar discharge sources and supersonic expansions have established that the rotational and vibrational motions of the He$_2$ molecules are not efficiently cooled in the expansions: rovibrational $a$ metastable states with $v^{\prime\prime}$ up to 2 and $N^{\prime\prime}$ up to 27 are significantly populated in the beam, corresponding to the distribution of population generated by the discharge. The abrupt ends of the rotational branches observed in Fig.~\ref{fig:FIG4_Rempi_a_c} thus do not reflect the distribution of population in the $a$ state. 
        Instead, we attribute them to tunneling predissociation through the barrier in the $c$ state potential (see Fig.~\ref{fig:FIG1_PEC_schemes}) on a time scale shorter than the 8~ns length of the laser pulses used experimentally. 
        In their previous study of these transitions by laser-induced-fluorescence (LIF) and He (1s)(2s) photofragment spectroscopy, Lorents {\it et al.} \cite{lorents89a} have observed the same phenomenon as a disappearance of the LIF signal coinciding with an increase He photofragment signal and a rapid broadening of the lines beyond $N^\prime = 12$.
        As expected from the relative positions of the $v^\prime=3$ and 4 levels of the $c$ state relative to the top of the barrier in the $c$ state potential, the onset of tunneling predissociation is observed at lower $N^\prime$ values for the $v^\prime = 4$ state ($N^\prime=12$) than is the case for the $v^\prime = 3$ state ($N^\prime=20$).
        
        The line positions observed in the present investigation agree within the experimental uncertainties with those determined in earlier spectroscopic studies \cite{ginter65a,lorents89a,kristensen90a} as well as in recent high-resolution measurements carried out in our laboratory \cite{wirth25a}.
        To maximize the population in the intermediate $c$ state when recording PFI-ZEKE PE spectra of the $X^+\leftarrow c$ ionizing transitions, both $v^\prime = 3$ and $v^\prime = 4$ vibrational levels were accessed from the $a(v^{\prime\prime} = 0)$ state, which is the most populated vibrational level in the supersonic beam.
        The broad range of rotational levels populated in the supersonic expansion is ideally suited to study high rovibrational levels of He$_2^+$ using the resonant excitation schemes described in Fig.~\ref{sec:experimental} (see Fig.~\ref{fig:FIG3_multiPanel_scheme_ZekeShift_Modified}). Moreover, the $v^\prime =3$ and 4 vibrational levels of the $c$ state have wavefunctions that extend to large internuclear distances. Consequently, they can be used to access high-lying vibrational levels of the $X^+$ state of He$_2^+$.

    \subsection{PFI-ZEKE photoelectron spectra of the electronic ground state of He$_2^+$}\label{subsec:PFIZEKEPE_Xp}
    
        Photoionization transitions to the $X^+$ ground state of He$_2^+$ from both the $a$ and $c$ states of He$_2$ were recorded using PFI-ZEKE PE spectroscopy.
        The electric-field pulse sequences used in these measurements yielded spectral lines with a full width at half maximum (FWHM) of $0.4\,\text{cm}^{-1}$, sufficient to fully resolve the rotational structure but not the spin-spin and spin-rotational fine-structure splittings. 
        Hund's angular-momentum-coupling case (b) was therefore used to analyze the rotational structure of the $a$ and $c$ states of He$_2$ and the $X^+$ ground state of He$_2^+$ and to derive the rovibrational photoionization selection rules. 
        
        In $^4$He$_2$ and $^4$He$_2^+$, the $I=0$ bosonic nature of the He$^{2+}$ nuclei implies that only even-$N$ (odd-$N$) rotational levels are populated in $\Sigma^+$ states of {\it gerade} ({\it ungerade}) electronic symmetry. The conservation of total angular momentum (neglecting spins) further leads to the photoionization selection rules \cite{xie90a,signorell97c}
        \begin{equation}\label{eq:Delta_N_Xa}
            \Delta N = N^+ - N^{\prime\prime} = 0, \pm 2, \ldots, \pm (l_{\rm pe} + 1), \quad \text{with $l_{\rm pe}$ odd}
        \end{equation}
        for the $X^+\leftarrow a$ photoionizing transition and 
        \begin{equation}\label{eq:Delta_N_Xc}
            \Delta N = N^+ - N^{\prime} = \pm 1, \pm 3, \ldots, \pm (l_{\rm pe} + 1), \quad \text{with $l_{\rm pe}$ even}
        \end{equation}
        for the $X^+\leftarrow c$ photoionizing transition, where $l_{\rm pe}$ is the orbital-angular-momentum quantum number of the photoelectron. 
        The $a$ state has the configuration $(1\sigma_g)^2(1\sigma_u)^1(2{\rm s}\sigma_g)^1$ so that single-photon ionization to the $X^+$ state, with configuration $(1\sigma_g)^2(1\sigma_u)^1$, implies the removal of the 2s ($l=0$) electron and the emission of a dominant $l_{\rm pe}=1$ partial wave. 
        Therefore, one expects the rotational structure of the $X^+\leftarrow a$ bands to consist of a dominant $\Delta N=0$ $Q$-Type rotational branch, with weaker $\Delta N=\pm 2$ O- and $S$-Type branches, as already discussed in Ref.~\cite{raunhardt08a}.
        The same arguments applied to the photoionization from the $c$ state, with dominant configuration $(1\sigma_g)^2(1\sigma_u)^1(3{\rm p}{\sigma_u})^1$ leads to the expectation of strong $\Delta N=\pm 1$ $P$- and $R$-Type branches in photoelectron spectra of the $X^+\leftarrow c$ transition under emission of $l_{\rm pe}=0,2$ photoelectron partial waves. 
        Weak $\Delta N=\pm 3$ $N$- and $T$-Type transitions might also be observed that reflect the electric quadrupole moment of the He$_2^+$ ion and a potential admixture of f character to the 3p$\sigma_u$ orbital of the $c$ state.

        PFI-ZEKE PE spectra of the lowest vibrational levels ($v^+=0-4$) of the $X^+$ state were recorded directly from the $a(v^{\prime\prime}=0-3)$ states populated in the supersonic expansion following single-photon ionization.
        Measurements of the $(0-0)$, $(1-1)$, and $(2-2)$ bands (not shown) confirmed the results obtained in Ref.~\cite{raunhardt08a} on the $v^+=0-2,N^+=1-27$ rovibrational levels of the $X^+$ state of He$_2^+$.
        Spectra of the $(3-2)$ and $(4-3)$ photoelectron bands enabled the determination of the positions of previously unobserved rotational levels of the $X^+(v^+ = 3,4)$ states. However, the low population of the $a(v^{\prime\prime}=2,3$) states limited the observations to only a few $Q$-Type transitions
        from rotational levels with $N^{\prime\prime} \leq 11$.
        The corresponding ionization energies are reported in \Cref{tab:results_measured_transitions}.

\renewcommand{\baselinestretch}{.76} 

        \begin{table*}[ht!]
            \centering
            \caption{
                Observed line positions (in cm$^{-1}$) of the $X^+~^2\Sigma_u^+(v^+, N^+ = N^{\prime\prime})\leftarrow a~^3\Sigma_u^+(v^{\prime\prime}, N^{\prime\prime})$ and $X^+~^2\Sigma_u^+(v^+, N^+ = N^{\prime} \pm 1)\leftarrow c~^3\Sigma_g^+(v^{\prime}, N^{\prime})$ ionizing transitions of He$_2$. 
                The transition values in parenthesis were not included in the final data set, the reason for their exclusion is given in the footnotes.
                The numbers in parentheses at the end of the transition wavenumbers represent the statistical uncertainties of the fitted central line positions, given in units of the last digit, and do not include the 0.1~cm$^{-1}$ systematic uncertainty. 
            }
            \scalebox{0.97}{ 
            \begin{ruledtabular}

				\begin{tabular}{ccc@{\hspace{-10pt}}cccccc}
                
                 & \multicolumn{1}{c}{$X^+(3)\leftarrow a(2)$} & $X^+(4)\leftarrow a(3)$ & &\multicolumn{2}{c}{$X^+(5)\leftarrow c(4)$} & \multicolumn{3}{c}{$X^+(6)\leftarrow c(4)$}\\
                $N^{\prime\prime}$ & \multicolumn{1}{c}{$Q$($N^{\prime\prime}$)} & $Q$($N^{\prime\prime}$) & $N^{\prime}$ & $R$($N^{\prime}$) & $P$($N^{\prime}$) & \multicolumn{2}{c}{$R$($N^{\prime}$)} & $P$($N^{\prime}$) \\\midrule
                1  &   \multicolumn{1}{c}{35588.619(20)} & (35424.59(5))\footnote{The line was too poorly resolved to be included in the fits.} & 0  &  25667.738(4) &       $-$        &  \multicolumn{2}{c}{{26943.125(11)}}&       $-$       \\
                3  &   \multicolumn{1}{c}{(35581.7(4))\footnote{The line overlaps with $X^+\leftarrow a$ transitions or autoionization lines, the uncertainty corresponds to the FWHM.}} & (35431.18(6))$^\textrm{a}$ & 2  &  25695.054(3) &   25635.660(4) &  \multicolumn{2}{c}{26968.003(9) }& 26911.014(7)  \\
                5  &   \multicolumn{1}{c}{35569.461(16)} & 35412.599(18)  & 4  &  25727.160(4) &   25620.585(5) &  \multicolumn{2}{c}{26995.772(13)}& 26893.597(8)  \\
                7  &   \multicolumn{1}{c}{35551.677(15)} & 35395.53(3)    & 6  &  25764.246(4) &   25610.870(8) &  \multicolumn{2}{c}{27026.599(11)}& 26879.528(10)  \\
                9  &   \multicolumn{1}{c}{35528.617(18)} & 35373.15(3)    & 8  &  25806.668(6) &   25607.115(4) &  \multicolumn{2}{c}{27060.777(11) }& 26869.659(10) \\
                11 &   \multicolumn{1}{c}{35500.258(21)} & 35345.793(22)  & 10 &  25854.971(5) &   25610.202(5) &  \multicolumn{2}{c}{27098.975(10)}& 26864.348(16) \\ \midrule
    
                & \multicolumn{2}{c}{$X^+(7)\leftarrow c(4)$} & \multicolumn{3}{c}{$X^+(8)\leftarrow c(4)$} & \multicolumn{3}{c}{$X^+(9)\leftarrow c(4)$}\\
                $N^{\prime}$ & $R$($N^{\prime}$) & $P$($N^{\prime}$) & \multicolumn{2}{c}{$R$($N^{\prime}$)} & $P$($N^{\prime}$) & \multicolumn{2}{c}{$R$($N^{\prime}$)} & $P$($N^{\prime}$) \\\midrule
                0  &  28147.298(16) &         $-$               &   \multicolumn{2}{c}{29280.398(21)} &        $-$       &  \multicolumn{2}{c}{30341.646(18)} & $-$  \\
                2  &  28169.856(9)  &   28115.350(8)            &   \multicolumn{2}{c}{29300.418(12)} &   29248.320(15) &  \multicolumn{2}{c}{30359.147(12)} & 30309.607(17)  \\
                4  &  28193.246(11) &   28095.358(14)           &   \multicolumn{2}{c}{29319.24(8)  } &   29225.892(20) &  \multicolumn{2}{c}{30373.45(3)  } & 30284.686(16)  \\
                6  &  28217.678(18) &   28076.966(14)           &   \multicolumn{2}{c}{29337.22(3)  } &   29202.94(4)   &  \multicolumn{2}{c}{30384.81(4)  } & 30257.17(6)    \\
                8  &  28243.565(20) &   28060.609(24)           &   \multicolumn{2}{c}{29354.58(8)  } &   29180.05(4)   &  \multicolumn{2}{c}{      $^\textrm{c}$    } &$^\textrm{c}$\\
                10 &  28271.44(3)   &   28047.18(4)             &   \multicolumn{2}{c}{    \footnote{The line was too weak to be detected.}       } &       $^\textrm{c}$    &  \multicolumn{2}{c}{    $^\textrm{c}$    } &$^\textrm{c}$\\ \midrule
    
                & \multicolumn{2}{c}{$X^+(10)\leftarrow c(4)$} & \multicolumn{3}{c}{$X^+(11)\leftarrow c(4)$}             & \multicolumn{3}{c}{$X^+(12)\leftarrow c(4)$}\\
                $N^{\prime}$ & $R$($N^{\prime}$)  & $P$($N^{\prime}$) & \multicolumn{2}{c}{$R$($N^{\prime}$)} & $P$($N^{\prime}$)       & \multicolumn{2}{c}{$R$($N^{\prime}$)} & $P$($N^{\prime}$) \\\midrule
                0            &  31330.752(22)   &       $-$       & \multicolumn{2}{c}{     \footnote{A problem with the calibration procedure prevented the determination of reliable transition wavenumbers.}     } &        $-$             & \multicolumn{2}{c}{ 33090.44(11)  } &        $-$      \\
                2            &  31345.624(7)    & 31298.687(11)   & \multicolumn{2}{c}{  32259.44(4)  } &     32215.186(16)     & \multicolumn{2}{c}{ 33099.981(20) } & 33058.40(3)  \\
                4            &  31355.306(11)   & 31271.070(15)   & \multicolumn{2}{c}{$^\textrm{d}$} &   $^\textrm{d}$                           & \multicolumn{2}{c}{ 33099.938(18) } & 33025.50(3)  \\
                6            &  31359.916(21)   & 31238.966(15)   & \multicolumn{2}{c}{       $^\textrm{d}$        } &        $^\textrm{d}$       & \multicolumn{2}{c}{ 33090.61(4)   } & 32983.52(7)  \\
                8            &  31359.886(19)   & 31202.804(17)   & \multicolumn{2}{c}{       $^\textrm{d}$        } &       $^\textrm{d}$        & \multicolumn{2}{c}{ 33072.11(5)   } & 32933.46(4)  \\
                10           & (31355.45(5))$^\textrm{a}$& 31163.38(4)    & \multicolumn{2}{c}{        $^\textrm{d}$       } &    $^\textrm{d}$             & \multicolumn{2}{c}{ 33045.12(6)  } & (32876.00(13))$^\textrm{a}$      \\ \midrule
    
                & \multicolumn{2}{c}{$X^+(10)\leftarrow c(3)$} & \multicolumn{3}{c}{$X^+(11)\leftarrow c(3)$} & \multicolumn{3}{c}{$X^+(12)\leftarrow c(3)$}\\
                $N^{\prime}$ & $R$($N^{\prime}$) & $P$($N^{\prime}$) & \multicolumn{2}{c}{$R$($N^{\prime}$)} & $P$($N^{\prime}$) & \multicolumn{2}{c}{$R$($N^{\prime}$)} & $P$($N^{\prime}$) \\\midrule
                10 &    $^\textrm{e}$        &   32202.02(3)     &  \multicolumn{2}{c}{    \footnote{This frequency range could not be scanned.}       }   &  33095.16(3)     & \multicolumn{2}{c}{       $^\textrm{e}$  } & $^\textrm{e}$ \\
                12 &  32360.85(8)   &   32134.50(4)     &  \multicolumn{2}{c}{33229.62(3)    }   &(33016.5(4))$^\textrm{b}$& \multicolumn{2}{c}{       $^\textrm{e}$  } & $^\textrm{e}$ \\
                14 &  32319.21(4)   &   32059.94(6)     &  \multicolumn{2}{c}{33172.04(3)    }   &  32928.35(5)     & \multicolumn{2}{c}{      $^\textrm{e}$    } & 33722.23(7) \\
                16 &  32269.75(3)   &      $^\textrm{b}$         &  \multicolumn{2}{c}{33104.75(3)    }   &  32832.32(3)     & \multicolumn{2}{c}{      $^\textrm{e}$   } & $^\textrm{c}$\\
                18 &  32213.439(20) &      $^\textrm{c}$         &  \multicolumn{2}{c}{33028.236(21)  }   &  32728.493(20)   & \multicolumn{2}{c}{      (33766.6(4))$^\textrm{b}$} & (33488.5(4))$^\textrm{b}$\\
                20 &      $^\textrm{c}$      &      $^\textrm{c}$         &  \multicolumn{2}{c}{(32943.82(4))$^\textrm{a}$}   &  32618.69(4)     & \multicolumn{2}{c}{  33658.56(3)} & $^\textrm{c}$ \\ \midrule
    
                & \multicolumn{2}{c}{$X^+(13)\leftarrow c(4)$} & \multicolumn{3}{c}{$X^+(14)\leftarrow c(4)$} & \multicolumn{3}{c}{$X^+(15)\leftarrow c(4)$}\\
                $N^{\prime}$ & $R$($N^{\prime}$) & $P$($N^{\prime}$) & \multicolumn{2}{c}{$R$($N^{\prime}$)} & $P$($N^{\prime}$)    & \multicolumn{2}{c}{$R$($N^{\prime}$)} & $P$($N^{\prime}$) \\\midrule
                0 &(33859.8(4))$^\textrm{b}$&        $-$              &  \multicolumn{2}{c}{  $^\textrm{b}$       }    &        $-$        &  \multicolumn{2}{c}{35172.90(4)  }& $-$\\
                2 &    $^\textrm{b}$       &(33827.5(4))$^\textrm{b}$          &  \multicolumn{2}{c}{34557.871(17)}    & (34522.5(4))$^\textrm{b}$     &  \multicolumn{2}{c}{35173.72(3)  }& 35140.92(4) \\
                4 &  33861.22(5)  &  33791.822(21)              &  \multicolumn{2}{c}{34547.647(14)}    &  34483.413(13)    &  \multicolumn{2}{c}{35158.102(21)}& 35099.35(5) \\
                6 &  33844.583(14)&  33745.03(6)                &  \multicolumn{2}{c}{(34523.6(4))$^\textrm{b}$}&  34431.318(24)    &  \multicolumn{2}{c}{35126.09(4)  }& $^\textrm{e}$\\
                8 &     $^\textrm{b}$      &      $^\textrm{b}$                   &  \multicolumn{2}{c}{34485.58(3)  }    &(34366.8(4))$^\textrm{b}$  &  \multicolumn{2}{c}{35077.84(4)  }& $^\textrm{e}$\\
                10 &     $^\textrm{e}$      &(33620.3(4))$^\textrm{b}$         &  \multicolumn{2}{c}{(34434.08(5))$^\textrm{a}$}&    $^\textrm{e}$           &  \multicolumn{2}{c}{$^\textrm{c}$}& $^\textrm{e}$\\\midrule
                
                & \multicolumn{2}{c}{$X^+(16)\leftarrow c(4)$} & \multicolumn{3}{c}{$X^+(17)\leftarrow c(4)$} & \multicolumn{3}{c}{$X^+(18)\leftarrow c(4)$}\\
                $N^{\prime}$ & $R$($N^{\prime}$) & $P$($N^{\prime}$) & \multicolumn{2}{c}{$R$($N^{\prime}$)} & $P$($N^{\prime}$) &\multicolumn{2}{c}{$R$($N^{\prime}$)} & $P$($N^{\prime}$) \\\midrule
                0  &   $^\textrm{b}$        &        $-$             &  \multicolumn{2}{c}{36179.76(4)}  &    $-$        & \multicolumn{2}{c}{(36565.9(4))$^\textrm{b}$}&        $-$       \\
                2  &   $^\textrm{b}$        &  35683.18(3)           &  \multicolumn{2}{c}{36174.19(3)}  & 36147.73(4)   & \multicolumn{2}{c}{36556.56(3)}  & 36533.531(21) \\
                4  & 35691.661(25)   &  35638.44(3)         &  \multicolumn{2}{c}{36147.07(4)}  & 36099.77(3)   & \multicolumn{2}{c}{36522.98(3)}  & 36482.05(3)   \\
                6  &   $^\textrm{b}$        &  35575.332(16)         &  \multicolumn{2}{c}{36098.05(4)}  & 36030.712(21) & \multicolumn{2}{c}{(36464.93(5))$^\textrm{a}$} & (36406.75(6))$^\textrm{a}$\\
                8  &  35592.348(20)&  35494.28(3)           &  \multicolumn{2}{c}{(36027.77(7))$^\textrm{a}$}  &  $^\textrm{b}$   & \multicolumn{2}{c}{(36382.03(5))$^\textrm{a}$} & $^\textrm{c}$ \\
                10 &  35514.73(5)  &    $^\textrm{e}$                &  \multicolumn{2}{c}{  $^\textrm{c}$     }  &     $^\textrm{e}$      & \multicolumn{2}{c}{$^\textrm{e}$} & $^\textrm{e}$\\\midrule
                & \multicolumn{2}{c}{ } & \multicolumn{3}{c}{$X^+(19)\leftarrow c(4)$} & \multicolumn{3}{c}{}\\
                $N^{\prime}$ & $R$($N^{\prime}$) & $P$($N^{\prime}$) & $N^{\prime}$ & \multicolumn{1}{c}{$R$($N^{\prime}$)} & $P$($N^{\prime}$) &$N^{\prime}$& $R$($N^{\prime}$) & $P$($N^{\prime}$) \\\midrule
                 0 &  $^\textrm{c}$ & $-$       &      2 & 36858.72(4) & 36839.46(4) & 4 & $^\textrm{c}$        & (36783.8(4))$^\textrm{b}$         \\
            \end{tabular}
            \end{ruledtabular}
            }
            \label{tab:results_measured_transitions}
        \end{table*}
     
        Employing the absorption-emission scheme illustrated in Fig.~\ref{fig:FIG1_PEC_schemes} enabled us to significantly enhance the population in the $a(v^{\prime\prime}=3-5$) vibrational levels through radiative decay of the $c(v^{\prime}=4$) levels.
        The branching ratios for the radiative decay of the c(4) level to these states, estimated to be 0.38 for $v^{\prime\prime}=3$, 0.02 for $v^{\prime\prime}=4$, and  0.26 for $v^{\prime\prime}=5$ from Franck-Condon factors calculated using the potential-energy functions reported by Yarkony \cite{yarkony89a}, were sufficient for the recording of good-quality PFI-ZEKE PE spectra of He$_2^+$ $X^+$ levels with $v^+$ up to 5, as demonstrated in Fig.~\ref{fig:FIG3_multiPanel_scheme_ZekeShift_Modified}e for the $X^+(v^+=5)\leftarrow a(v^{\prime\prime}=5)$ transition. Reaching higher vibrational levels of the $X^+$ state of He$_2^+$ with this scheme would require successive absorption-emission cycles and turned out to be impractical both in terms of the signal intensity and the number of lasers required.
        
\begin{figure*}[ht]
            \begin{center}
                \includegraphics[width=\linewidth]{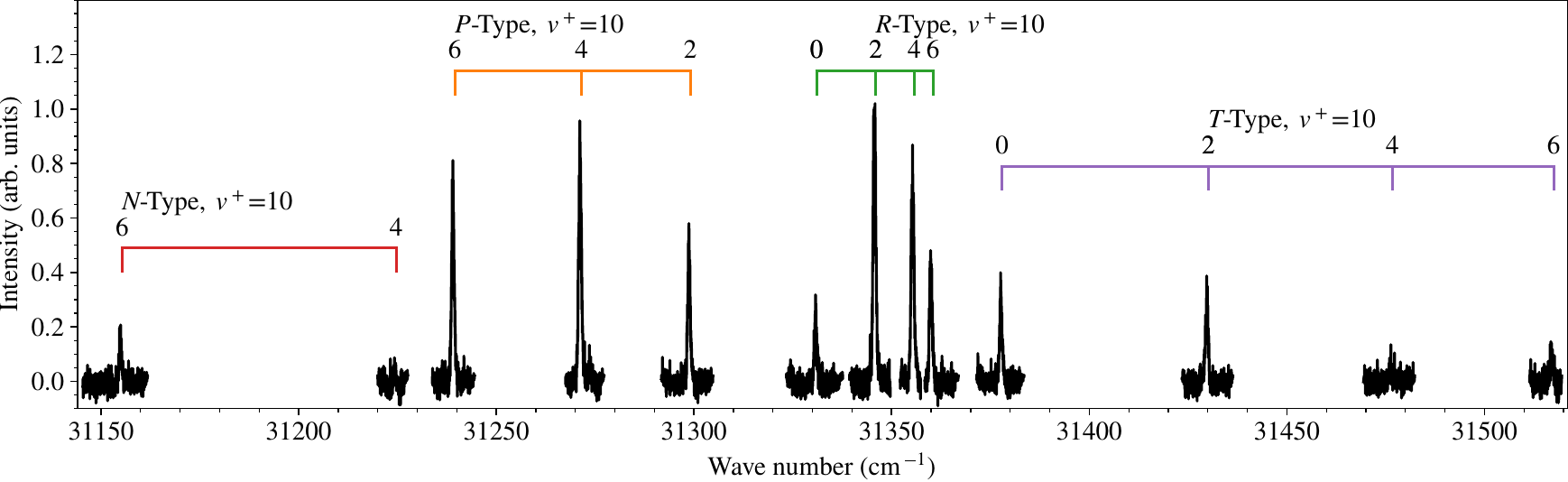}
                \caption{\label{fig:FIGExtra_N_T_Branches}
                PFI-ZEKE PE spectra of the $\Delta N = -3, -1, 1$ and 3 rotational branches (labeled $N$-, $P$-, $R$- and $T$-Type branches, respectively) of the $X^+~^2\Sigma_u^+(v^+=10)\leftarrow c~^3\Sigma_g^+(v^\prime=4)$ transition recorded in disjoint segments. The values of the rotational quantum number of the selected $c$ state intermediate level is given along the assignment bars.}
            \end{center}
        \end{figure*}

        A much more efficient approach to study vibrationally excited levels of the $X^+$ state of He$_2^+$ turned out to be by direct photoionization from the intermediate $c(v^{\prime}=3,4; N^\prime)$ states. Their strongly delocalized (tunneling) vibrational wavefunctions provided access to $X^+$ states with $v^+$ up to 19. Ionization from the $c$ state of He$_2$ to form the $X^+$ electronic ground state is expected to lead to the emission of s ($l_{\rm pe}=0$) and d ($l_{\rm pe}=2$) photoelectron partial waves, which allows the observation of transitions with $\Delta N = \pm 1, \pm 3, \ldots$ [see Eq.~(\ref{eq:Delta_N_Xc}) and discussion above]. All four branches were observed experimentally, as illustrated in Fig.~\ref{fig:FIGExtra_N_T_Branches} with the example of the  $X^+(10,N^+)\leftarrow c (4,N^\prime)$ transitions. This spectrum consists of dominant $P$-, and $R$-Type branches, with weaker $N$- and $T$-Type transitions, as expected upon removal of the 3p electron from the $c$ state configuration. 
    
        \begin{figure*}[ht]
            \begin{center}
                \includegraphics[width=\linewidth]{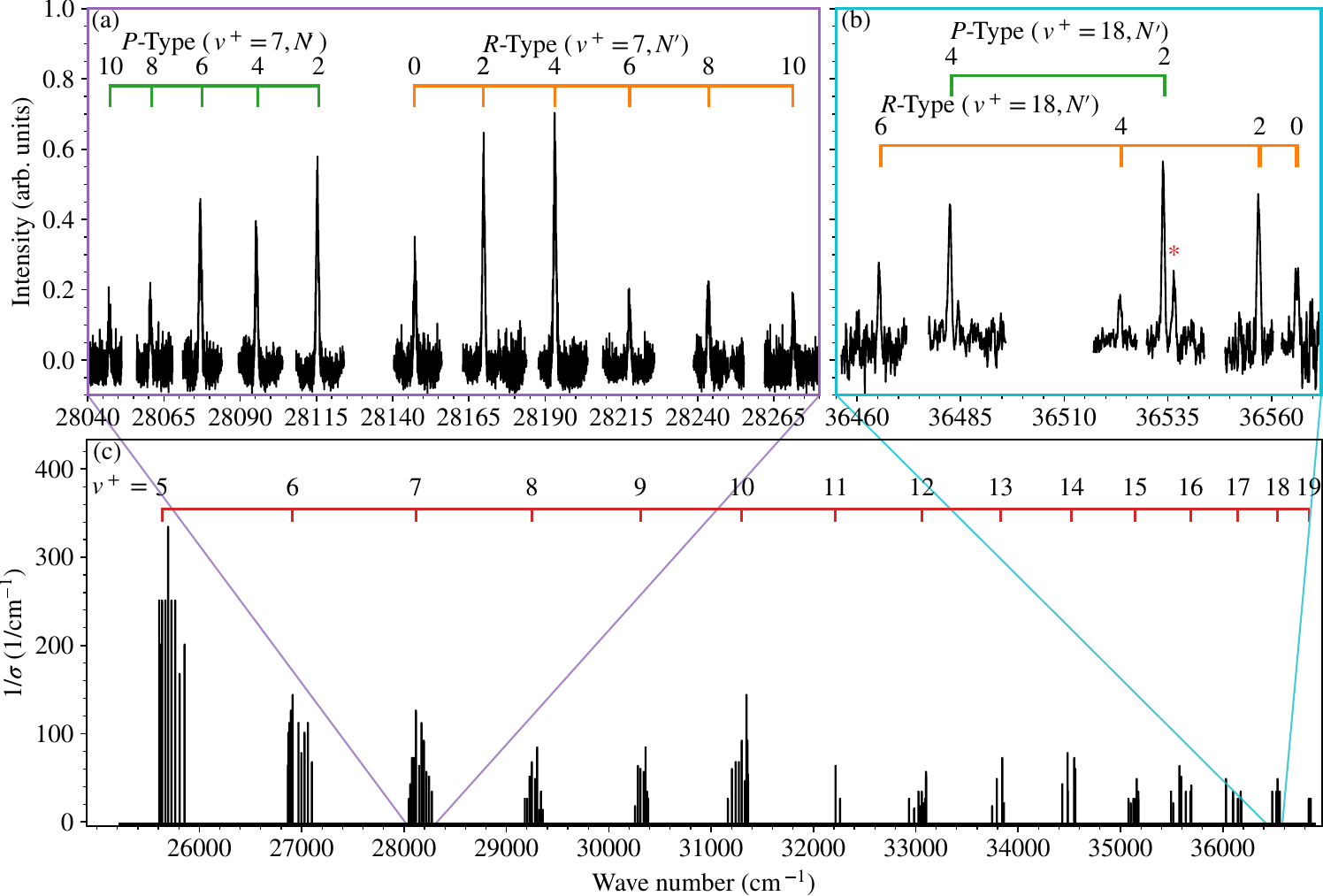}
                \caption{\label{fig:FIG5_overviewZeke_insets}
                (a) and (b) PFI-ZEKE PE spectra of the $P$- and $R$-Type branches of the $X^+~^2\Sigma_u^+(v^+=6\ {\rm and}\ 18, N^+)\leftarrow  c~^3\Sigma_g^+(v^\prime=4, N^\prime)$ transitions, respectively, recorded in disjoint segments. 
                The red asterisk indicates a line that did not arise from transitions from the $c~^3\Sigma_g^+(v^\prime=4)$ state but either from an ionizing transition from the $a~^3\Sigma_u^+$ state or from an autoionization process. (c) Stick spectrum indicating the positions of all measured $X^+~^2\Sigma_u^+(v^+, N^+)\leftarrow c~^3\Sigma_g^+(v^\prime=4, N^\prime)$ ionizing transitions of He$_2$. 
                The heights of the sticks correspond to the inverse of the statistical uncertainty in the respective fitted line-center positions.
                The assignment bars label the final vibrational states of the rotational ($P$- or $R$-Type) transitions connecting the $c~^3\Sigma_g^+(v^\prime=4)$ to the $X^+~^2\Sigma_u^+(v^+)$ states.
                }
            \end{center}
        \end{figure*}
        The procedure followed to systematically determine the positions of the different rotational levels of the vibrationally excited $X^+(v^+\ge 5)$ states is illustrated in Fig.~\ref{fig:FIG5_overviewZeke_insets}a and b with the examples of spectra of the $P$- and $R$-Type branches of the $X^+(7)\leftarrow c(4)$ and $X^+(18)\leftarrow c(4)$ transitions, respectively. 
        The former spectrum consists of 11 lines, corresponding to transitions to $X^+(v^+=7,N^+=1-11)$ levels, recorded in different scans after selectively exciting one of the $c(v=4,N^\prime =0,2,4,6,8$ or $ 10)$ levels using $P$(1), $R$(1), $R$(3), $R$(5), $R$(7), and $R$(9) lines of the $c(4)\leftarrow a(0)$ band. 
        The latter spectrum ($v^+=18$), recorded at an increased laser intensity using the same procedure, is much weaker, reflecting the decrease of the Franck-Condon factors with increasing $v^+$ value. 
        It consists of transitions from selected $c(4)$ intermediate levels with $N^\prime=0,2,4$ and 6 to $X^+(18,N^+=1,3,5$ and 7) levels as well as weaker transitions originating from other processes, primarily photoionizing transitions from rovibrationally excited $a$ state levels. 
        The horizontal scale in these spectra represents the wavenumbers of the $X^+(v^+,N^+)\leftarrow c(4,N^\prime)$ ionization thresholds after correction of the field-induced shifts of the ionization thresholds, as explained in Section~\ref{sec:experimental} (see  Fig.~\ref{fig:FIG3_multiPanel_scheme_ZekeShift_Modified}d and e).
        
        The very different behavior of the $P$-Type and $R$-Type rotational branches of the $X^+(v^+=7\ {\rm and }\ 18)\leftarrow c(4)$ photoelectron bands depicted in Fig.~\ref{fig:FIG5_overviewZeke_insets} is a consequence of the large reduction of the rotational constants of the $X^+(v^+)$ states as $v^+$ increases from 7 ($B_7^+= 5.40$ cm$^{-1}$) to 18 ($B_{18}^+= 2.28$ cm$^{-1}$). 
        At $v^+=7$, the rotational constant of He$_2^+$ is slightly larger than the $c(v^\prime=4)$ rotational constant ($B^\prime_4 = 5.34$ cm$^{-1}$) and, consequently, the spacings between adjacent $P$-Type ($R$-Type) branches slightly decrease (increase) with increasing $N^\prime$ value. 
        At $v^+=18$, the rotational constant of He$_2^+$ is so small that the $P$- and $R$-Type branches both evolve toward lower wavenumbers as $N^\prime$ increases. 
        
        No photoelectron spectra could be recorded from $c(v^\prime =4,N^\prime>10$) levels because of the rapid decay of these levels by tunneling predissociation, which limited the measurements of PFI-ZEKE PE spectra to $X^+(v^+,N^+\leq11)$ levels (see Fig.~\ref{fig:FIG4_Rempi_a_c}). 
        To access higher rotational levels of the $X^+$ state for $v^+$ values in the range $10-13$, $c(v^\prime = 3, 10\leq N^\prime\leq 20)$ levels were used as intermediate states. These states are either stable or metastable with respect to predissociation and their vibrational wavefunctions enable access to ionic levels with $v^+\geq 10$. 
        In the case of the $v^+=11$ level, the use of $c(3)$ intermediate levels turned out to be particularly useful because the overlap of the $X^+(11)\leftarrow c(4)$ band with the much stronger $X^+(0-2)\leftarrow  a (0-2)$ bands prevented the observation of almost all lines of the $X^+(11)\leftarrow c(4)$ band.
        
        An overview of all $X^+(v^+,N^+)\leftarrow c(4,N^\prime)$ lines that were measured in the present work is presented in Fig.~\ref{fig:FIG5_overviewZeke_insets}c. 
        For clarity, the lines are presented as sticks at the positions of the corresponding field-corrected ionization thresholds and their heights correspond to the inverse standard deviations resulting from weighted fits of the line profiles to Gaussian functions. 
        The $v^+=5$ ionization thresholds could be determined with a relative uncertainty of better than 0.005~cm$^{-1}$ because of the high signal-to-noise ratio of the corresponding spectra. In contrast, the ionization thresholds corresponding to the highest vibrational levels could only be determined with relative uncertainties of about 0.05~cm$^{-1}$ because of the much poorer signal-to-noise ratios (see Fig.~\ref{fig:FIG5_overviewZeke_insets} for the transitions to $X^+(v^+=18)$ levels).
        Because of accidental overlaps of the lines of the $X^+-c$ and $X^+-a$ band systems, several lines could not be observed. 
        The regions 32\,300$-$32\,800~cm$^{-1}$ and 34\,000$-$34\,700~cm$^{-1}$, where the $\Delta v=-1$ and $\Delta v=0$ bands of the $X^+(v^+)\leftarrow a(0-2)$ transitions pile up, turned out to be particularly challenging. 
        
        The positions of all ionization thresholds determined in the present work are listed in Table~\ref{tab:results_measured_transitions}, where the numbers in parentheses represent the statistical uncertainties ($1\sigma$) in the determination of the line centers and do not include the systematic uncertainty of $\approx 0.1$~cm$^{-1}$ resulting from the calibration procedure (see Section~\ref{sec:experimental}). 
        These data considerably extend the experimental information available on the rovibrational level structure of the $X^+$ state of He$_2^+$ and almost completely bridge the gap between the low-$v^+$ region ($v^+=1-3$), which was previously studied by high-resolution photoelectron spectroscopy \cite{raunhardt08a,motsch14a} and Rydberg extrapolation methods \cite{ginter84a,sprecher14a,semeria16a,jansen16a,jansen18a,jansen18b,semeria20a}, and the region of the highest vibrational levels ($v^+=22,23$), which was studied by microwave electronic spectroscopy \cite{carrington95b}.
        
    \subsection{Rovibrational level structure of the $X^+$ $^2\Sigma_u^+$ state of He$_2^+$ }\label{subsec:comparisonMeasuredCalculated}

        The $X^+(v^+,N^+)\leftarrow a(v^{\prime\prime},N^{\prime\prime})$ and $X^+(v^+,N^+)\leftarrow c(v^{\prime},N^{\prime})$ ionization thresholds listed in Table~\ref{tab:results_measured_transitions} were used to determine the rovibrational level structure of the $X^+$ electronic ground state of He$_2^+$. 
        In order to derive a complete and accurate description of this level structure, we combined the results summarized in Table~\ref{tab:results_measured_transitions} with the results obtained in earlier studies of the $X^+(v^+=0, 1$ and 2) vibrational levels by high-resolution photoelectron and photoionization spectroscopy \cite{raunhardt08a,sprecher14a,jansen16a,jansen18b,semeria20a}. 
        We followed a general procedure, described in detail by Albritton {\it et al.} \cite{albritton76a}, and extracted the level structure in a weighted least-squares fit from a redundant set of measured transitions and ionization wavenumbers connecting the rovibrational levels of the $a$, $c$ and $e$ $^3\Pi_g$ states of He$_2$ and the $X^+$ state of He$_2^+$. 
        In addition to the ionization wavenumbers listed in Table~\ref{tab:results_measured_transitions}, we included the $c(v^{\prime}=0-2,N^{\prime})\leftarrow a(v^{\prime\prime}=0-2,N^{\prime\prime}$) transition wavenumbers reported by Focsa {\it et al.} \cite{focsa98a} and the $e(v^{\prime}=0-4,N^{\prime})\leftarrow a(v^{\prime\prime}=0-5,N^{\prime\prime})$ transition wavenumbers determined by Brown {\it et al.} \cite{brown71a}. The inclusion of these transition wavenumbers enabled all ionic levels to be embedded in a single network of transitions and ensured a reliable and accurate description of the relative positions of the $c (v=3,4)$ and $a (v=0-5)$ states that were used to reach the $X^+(v^+, N^+)$ ionization thresholds in the present work. 
        
        The complete list of transitions used in the linear least-squares fit and the fitted rovibrational term values are provided in Tables S1 and S2 of the supplemental material. 
        The rotational levels of the $X^+$ state of He$_2^+$ relative to the centroid position of the $X^+(v^+=0,N^+=1)$ ground state are listed with their experimental uncertainties in Table~\ref{tab:energy_levels_Xp}, where they are compared with the term values obtained by M\'atyus and coworkers in calculations including relativistic and quantum-electrodynamic corrections \cite{matyus25a}. 
        These calculations, carried out in parallel to the present work, improved the results of an earlier set of calculations based on a less accurate Born-Oppenheimer potential-energy function \cite{matyus18a}. 
        The residuals [${\rm experimental} - {\rm calculated}$] corresponding to these two sets of calculations are displayed in Fig.~\ref{fig:FIG6_PEC_e-c_comparison}b) and (c) and reveal a remarkable agreement. 
        Whereas the residuals of the earlier calculations exhibit slight systematic deviations that increase with increasing degree of rovibrational excitation (Fig.~\ref{fig:FIG6_PEC_e-c_comparison}b), reaching values of $\sim 0.6$ cm$^{-1}$ for the highest vibrational levels observed experimentally ($v^+=16-19$), the results of the most recent calculations agree with the experimental results within the experimental uncertainties of 0.1 cm$^{-1}$ over the entire range of observed rovibrational levels, from the ($v^+=0,N^+=1$) ground state (rovibrational term value of 0~cm$^{-1}$) to the $(v^+=19,N^+=3)$ level with a rovibrational term value of 18\,659.00(20)~cm$^{-1}$ (Fig.~\ref{fig:FIG6_PEC_e-c_comparison}c).

\setlength{\tabcolsep}{4.4pt}
\renewcommand{\baselinestretch}{1.1} 

        \begin{table*}[ht!]
            \caption{Experimental term values $T_{v^+\,N^+}$ (in cm$^{-1}$) of the rovibrational states of the $X^+$ $^2\Sigma_u^+$ state of He$_2$ determined in the present work with respect to the $X^+$ $^2\Sigma_u^+$($v^+=0, N^+=1$) ground state. 
            The numbers in parentheses indicate the estimated uncertainty in units of the last digit.
            The columns labeled $\Delta^\textrm{a}$ and $\Delta^\textrm{b}$ list the differences (${\rm experimental} - {\rm calculated}$, also in cm$^{-1}$) between the measured term values and those calculated \textit{ ab initio} by M\'atyus {\it et al.} \cite{matyus25a} and those obtained from the effective potential-energy function $V_{\rm eff}(R)$ with the potential parameters listed in Table~\ref{tab:potential}, respectively. See text for details.
            }\label{tab:energy_levels_Xp}
            \centering
            \scalebox{0.98}{ 

            \begin{ruledtabular}

            \begin{tabular}{ccc}
                \begin{tabular}[t]{lccc}
                    $v^+, N^+$ & $T_{v^+\,N^+}$ & $\Delta$\footnote{Levels calculated by M\'atyus {\it et al.} \cite{matyus25a}.} & $\Delta$\footnote{Levels calculated using the effective potential determined in this work.} \\\midrule
                    0,1 &      0.000(3) & 0.000 & 0.000 \\
                    0,3 &     70.938(4) & 0.000 & 0.000 \\
                    0,5 &    198.365(4) & 0.001 & 0.001 \\
                    0,7 &    381.835(4) & 0.003 & 0.003 \\
                    0,9 &    620.702(4) & 0.002 & 0.001 \\
                    0,11 &    914.137(4) & 0.001 & 0.000 \\
                    0,13 &   1261.124(4) & 0.001 & 0.000 \\
                    0,15 &   1660.463(4) & $-$0.002 & $-$0.002 \\
                    0,17 &   2110.793(4) & 0.000 & 0.000 \\
                    0,19 &   2610.574(4) & $-$0.003 & 0.000 \\
                    0,21 &   3158.16(7) & 0.03 & 0.04 \\
                    0,23 &   3751.67(11) & 0.06 & 0.07 \\
                    1,1 &   1627.932(5) & 0.000 & 0.001 \\
                    1,3 &   1696.610(5) & 0.000 & 0.000 \\
                    1,5 &   1819.90(7) & $-$0.07 & $-$0.07 \\
                    1,7 &   1997.563(5) & 0.000 & 0.001 \\
                    1,9 &   2228.70(10) & 0.00 & 0.00 \\
                    1,11 &   2512.687(5) & $-$0.002 & $-$0.001 \\
                    1,13 &   2848.369(5) & $-$0.002 & 0.001 \\
                    1,15 &   3234.60(10) & 0.00 & 0.00 \\
                    1,17 &   3669.90(10) & $-$0.10 & $-$0.10 \\
                    2,1 &   3185.53(11) & $-$0.08 & $-$0.06 \\
                    2,3 &   3252.00(10) & 0.00 & 0.00 \\
                    2,5 &   3371.30(10) & 0.10 & 0.10 \\
                    2,7 &   3543.00(10) & 0.10 & 0.10 \\
                    2,9 &   3766.30(10) & 0.00 & 0.00 \\
                    3,1 &   4672.93(7) & 0.00 & 0.03 \\
                    3,5 &   4852.12(6) & 0.02 & 0.05 \\
                    3,7 &   5017.71(5) & $-$0.03 & 0.01 \\
                    3,9 &   5233.26(6) & $-$0.02 & 0.01 \\
                    3,11 &   5497.88(7) & $-$0.02 & 0.01 \\
                    4,5 &   6262.25(7) & $-$0.14 & $-$0.10 \\
                    4,7 &   6421.89(11) & $-$0.05 & $-$0.01 \\
                    4,9 &   6629.45(11) & $-$0.07 & $-$0.03 \\
                    4,11 &   6884.22(8) & $-$0.07 & $-$0.03 \\
                    5,1 &   7435.944(22) & $-$0.060 & 0.000 \\
                    5,3 &   7495.330(20) & $-$0.040 & 0.020 \\
                    5,5 &   7601.908(21) & $-$0.061 & $-$0.006 \\
                \end{tabular}

                \begin{tabular}[t]{lccc}
                    $v^+, N^+$ & $T_{v^+\,N^+}$ & $\Delta^\textrm{a}$ & $\Delta^\textrm{b}$ \\\midrule
                    5,7 &   7755.283(19) & $-$0.060 & $-$0.009 \\
                    5,9 &   7954.826(22) & $-$0.017 & 0.029 \\
                    5,11 &   8199.59(3) & $-$0.04 & 0.01 \\
                    6,1 &   8711.31(3) & $-$0.09 & $-$0.02 \\
                    6,3 &   8768.31(3) & $-$0.05 & 0.01 \\
                    6,5 &   8870.55(3) & $-$0.07 & $-$0.01 \\
                    6,7 &   9017.74(3) & 0.01 & 0.06 \\
                    6,9 &   9208.95(3) & $-$0.08 & $-$0.04 \\
                    6,11 &   9443.60(6) & $-$0.08 & $-$0.04 \\
                    7,1 &   9915.60(3) & $-$0.12 & $-$0.06 \\
                    7,3 &   9970.12(3) & $-$0.11 & $-$0.06 \\
                    7,5 &  10068.00(3) & $-$0.09 & $-$0.04 \\
                    7,7 &  10208.74(5) & $-$0.10 & $-$0.05 \\
                    7,9 &  10391.74(6) & $-$0.07 & $-$0.03 \\
                    7,11 &  10616.06(11) & $-$0.10 & $-$0.06 \\
                    8,1 &  11048.60(5) & $-$0.07 & $-$0.02 \\
                    8,3 &  11100.68(4) & $-$0.02 & 0.03 \\
                    8,5 &  11193.99(12) & $-$0.09 & $-$0.05 \\
                    8,7 &  11328.25(8) & $-$0.10 & $-$0.07 \\
                    8,9 &  11502.7(3) & $-$0.1 & $-$0.1 \\
                    9,1 &  12109.87(5) & $-$0.03 & 0.01 \\
                    9,3 &  12159.43(4) & 0.03 & 0.07 \\
                    9,5 &  12248.20(9) & $-$0.01 & 0.02 \\
                    9,7 &  12375.86(14) & $-$0.03 & 0.00 \\
                    10,1 &  13098.96(4) & $-$0.02 & 0.01 \\
                    10,3 &  13145.89(3) & 0.00 & 0.03 \\
                    10,5 &  13230.04(3) & $-$0.01 & 0.01 \\
                    10,7 &  13350.96(5) & $-$0.02 & $-$0.01 \\
                    10,9 &  13508.03(5) & 0.03 & 0.03 \\
                    10,11 &  13700.16(14) & $-$0.06 & $-$0.07 \\
                    10,13 &  13926.45(17) & $-$0.08 & $-$0.10 \\
                    10,15 &  14185.69(14) & 0.05 & 0.02 \\
                    10,17 &  14476.10(11) & 0.06 & 0.02 \\
                    10,19 &  14796.08(7) & 0.08 & 0.02 \\
                    11,1 &  14015.46(6) & 0.07 & 0.07 \\
                    11,3 &  14059.72(14) & 0.06 & 0.07 \\
                    11,9 &  14401.15(11) & 0.09 & 0.07 \\
                    11,13 &  14795.16(9) & 0.18 & 0.14 \\
                \end{tabular}

                \begin{tabular}[t]{lccc}
                    $v^+, N^+$ & $T_{v^+\,N^+}$ & $\Delta^\textrm{a}$ & $\Delta^\textrm{b}$ \\\midrule
                    11,15 &  15038.59(7) & 0.13 & 0.08 \\
                    11,17 &  15311.12(6) & 0.16 & 0.10 \\
                    11,19 &  15610.85(7) & 0.15 & 0.07 \\
                    12,1 &  14858.70(10) & 0.10 & 0.10 \\
                    12,3 &  14900.25(6) & 0.14 & 0.12 \\
                    12,5 &  14974.68(6) & 0.07 & 0.05 \\
                    12,7 &  15081.60(10) & 0.00 & 0.00 \\
                    12,9 &  15220.27(17) & 0.03 & $-$0.01 \\
                    12,11 &  15389.74(21) & 0.02 & $-$0.04 \\
                    12,13 &  15588.71(24) & $-$0.13 & $-$0.20 \\
                    12,21 &  16650.61(11) & 0.03 & $-$0.09 \\
                    13,3 &  15666.57(7) & 0.03 & $-$0.02 \\
                    13,5 &  15736.01(13) & 0.01 & $-$0.04 \\
                    13,7 &  15835.63(5) & $-$0.02 & $-$0.08 \\
                    14,3 &  16358.16(4) & 0.00 & $-$0.06 \\
                    14,5 &  16422.39(4) & $-$0.02 & $-$0.09 \\
                    14,9 &  16633.74(11) & 0.04 & $-$0.04 \\
                    15,1 &  16941.20(10) & 0.00 & 0.00 \\
                    15,3 &  16974.02(9) & $-$0.02 & $-$0.08 \\
                    15,5 &  17032.85(7) & $-$0.04 & $-$0.10 \\
                    15,7 &  17117.14(14) & $-$0.01 & $-$0.07 \\
                    15,9 &  17226.00(14) & $-$0.01 & $-$0.07 \\
                    16,1 &  17483.46(11) & 0.02 & 0.00 \\
                    16,3 &  17513.19(11) & 0.01 & $-$0.02 \\
                    16,5 &  17566.39(5) & 0.01 & $-$0.02 \\
                    16,7 &  17642.44(11) & 0.00 & $-$0.02 \\
                    16,9 &  17740.51(7) & 0.01 & 0.00 \\
                    16,11 &  17859.35(17) & $-$0.04 & $-$0.06 \\
                    17,1 &  17948.00(10) & 0.00 & 0.00 \\
                    17,3 &  17974.49(8) & 0.01 & 0.04 \\
                    17,5 &  18021.77(7) & 0.03 & 0.06 \\
                    17,7 &  18089.10(14) & $-$0.06 & $-$0.03 \\
                    18,1 &  18333.81(8) & 0.04 & 0.08 \\
                    18,3 &  18356.82(8) & 0.07 & 0.11 \\
                    18,5 &  18397.73(11) & 0.01 & 0.05 \\
                    19,1 &  18639.74(14) & 0.10 & 0.07 \\
                    19,3 &  18659.00(14) & 0.10 & 0.07 
                \end{tabular}
            \end{tabular}
            \end{ruledtabular}
        }
            \label{tab:results_energy_levels}
        \end{table*}     

        \subsection{Empirical effective potential-energy function of the $X^+$ $^2\Sigma_u^+$ state of He$_2^+$}\label{subsec:potential}

		\begin{figure*}[ht]
            \begin{center}
                \includegraphics[width=0.85\textwidth]{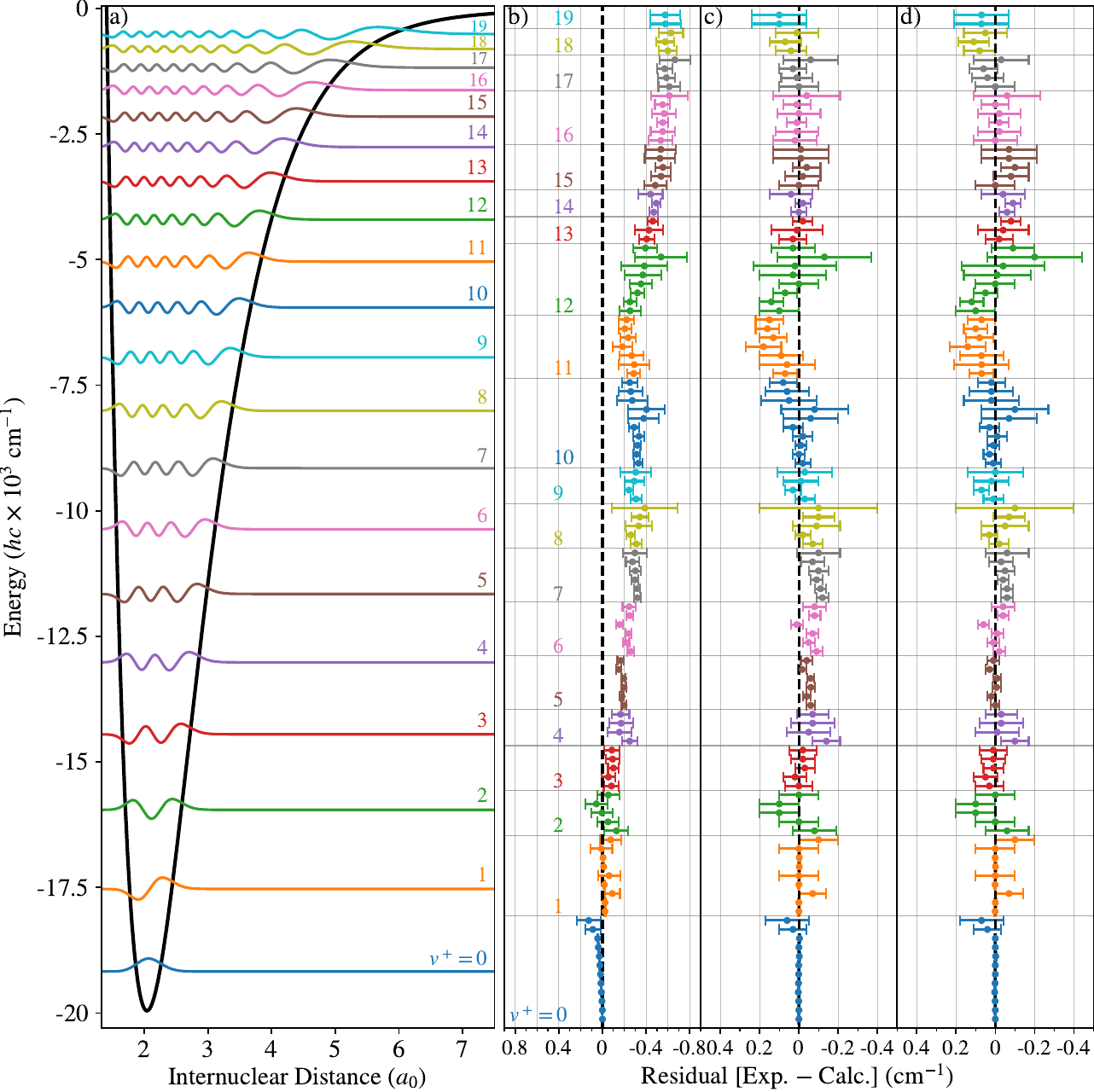}
                \caption{\label{fig:FIG6_PEC_e-c_comparison}
                (a) Energy-level diagram showing the positions and the wavefunctions of all $v^+, N^+=1$ vibrational levels of the $X^+$ $^2\Sigma_u^+(v^+, N^+)$ electronic ground state of He$_2^+$ calculated from the potential-energy function obtained in the present work.
                (b)-(d) Differences [${\rm experimental} - {\rm calculated}$] of the term values of the $X^+$ $^2\Sigma_u^+(v^+, N^+)$ levels of He$_2^+$ determined experimentally in the present work and these calculated \textit{ab initio} by M\'atyus in 2018 \cite{matyus18a} (b) and in 2025 after improving the theoretical treatment \cite{matyus25a} (c) and those obtained from the empirical effective potential-energy function determined in the present work (d).
                }
            \end{center}
        \end{figure*}

        The experimental data on the energy-level structure of the $X^+$ ground state of He$_2^+$ consist of four disjoint sets: (i) Rovibrational energies in the ground state of $^4$He$_2^+$ with vibrational and rotational quantum numbers in the ranges between $v^+=0$ and 19 and $N^+=1$ and 27, respectively, as presented in the previous section; (ii) the frequencies, in the microwave range, of seven transitions involving the highest bound levels [($v^+,N^+)=(22,5),\ (23,1)$ and (23,3) ] of the $X^+$ electronic ground state and the weakly bound levels (0,0), (0,2), (0,4), (1,0) and (1,2) of the $A^+$ $^2\Sigma_g^+$ first electronically excited state \cite{carrington95b}, (iii) the frequencies, in the IR, of nine $R$-Type branch transitions connecting the rotational levels $N^+=1,3-11$ of the $v^+=0$ ground vibrational level of $^4$He$^3$He$^+$ to the corresponding $N^++1$ rotational levels of the $v^+=1$ first vibrationally excited level reported by Yu and Wing \cite{yu87a,yu88b}, and (iv) the energies of the $N^+=0-11$ rotational levels in the vibrational ground state of $^3$He$_2^+$ derived from the photoelectron spectrum of $^3$He$_2$ by Raunhardt {\it et al.} \cite{raunhardt08a}. 
        This section is devoted to the determination of an “effective” potential-energy function for the electronic ground state of He$_2^+$ that reproduces these experimental results. 

		\begin{table*}[ht!]
            \centering
            \caption{Values of the parameters and coefficients of the effective potential of the $X^+$ $^2\Sigma_u^+$ state of He$_2^+$ obtained in a least-squares fit to the experimental data.}
            
            \begin{ruledtabular}

                \begin{tabular}{ccclcccc}
                    {Param.} & {Value} &  {Param.} & {Value} & {$\sigma$}\footnote{Using the notation $x[y]=x\times 10^y$} & {Param.} & {Value} & {$\sigma$}$^\textrm{a}$  \\
                    \midrule
                    $R_\textrm{m}$    & 2.042                                            $a_0$ &  $b$     & 7.57773944$\times 10 ^ {-1}$   & 9[-3]   & $K_{10}$  & $-4.70535910 \times 10 ^ 6$ & 1.5[5]\\
                    $R_\textrm{c}$    & 6.15                                             $a_0$ &  $a_{0}$ & $-9.09266603\times 10 ^ {-2}$ & 1.3[-7] & $K_{11}$  & $3.29238026 \times 10 ^ 8$  & 9[6]\\
                    $\varGamma$          & 0.05                                             $a_0$ &  $a_{1}$ & $-1.41519583\times 10 ^ {-4}$  & 1.6[-6] & $K_{12}$  & $-9.30812659\times 10 ^ 9$ & 2.5[8]\\
                    $R_\textrm{e}$    & 2.042\,182\,22                                   $a_0$ &  $a_{2}$ & 1.40824402      & 1.5[-2] & $K_{13}$  & $1.33640532 \times 10 ^ {11}$ & 3[9]\\
                    $D_\textrm{e}$    & 19\,956.10(10)                               cm$^{-1}$  &  $a_{3}$ & $-2.66855326$     & 6[-2]   & $K_{14}$  & $-9.63352030\times 10 ^{11}$& 3[10]\\
                    $C_{4}$           & 0.691\,596\,087\,20\footnote{Computed following G\c{e}bala \textit{et al.}~\cite{gebala23a} using values reported in Refs. \cite{davidson66, bishop89a, yan96a, yan98a}.}                                &  $a_{4}$ & 3.64183818$\times 10 ^ {-1}$   & 5[-2]   & $K_{15}$  & $1.95122170 \times 10 ^{12}$& 2.1[11]\\
                    $C_{6}$           & 1.5968$^\textrm{b}$                                            &  $a_{5}$ & 1.30030450      & 3[-2]   & $K_{16}$  & $2.05596750 \times 10 ^{13}$& 1.5[12]\\
                    $C_{7}$           & 3.664$^\textrm{b}$                                            &  $a_{6}$ & 1.01688778$\times 10 ^ {-1}$   & 5[-2]   & $K_{17}$  & $-1.69997224 \times 10 ^{14}$& 8[12]\\
                    $C_{8}$           & 9.818$^\textrm{b}$                                              &  $a_{7}$ & $-2.38312184$     & 1.5[-1] & $K_{18}$  & $5.16523633 \times 10 ^{14}$ & 2.2[13]\\
                    $C_{9}$           & 37.84$^\textrm{b}$                                              &  $a_{8}$ & 4.31311306      & 2.3[-1] & $K_{19}$  & $-5.92029761 \times 10 ^{14}$& 2.5[13]\\
                                      &                                                        &  $a_{9}$ & $-4.13851208$     & 1.3[-1] &           &                & \\
                            \end{tabular}
                \end{ruledtabular}

            \label{tab:potential}
        \end{table*}
        
        Previous theoretical work and {\it ab initio} calculations have quantified nonadiabatic, relativistic and quantum-electrodynamics corrections to the rovibrational levels of the $X^+$ ground state of He$_2^+$ \cite{xie05a,pachucki06a,tung12a,matyus18a,matyus25a}. Although these corrections go beyond the solution of the Schrödinger equation of the nuclear motion based on a single adiabatic potential, they might nevertheless be accurately accounted for in an effective manner in an adiabatic calculation based on a single potential-energy function. Such a function would not only be helpful to fill the gaps between the four experimental data sets listed above; it would also allow the straightforward calculation of the shape resonances of He$_2^+$. We follow the approach of Tung {\it et al.} \cite{tung12a}, who demonstrated that nonadiabatic corrections can be included in an effective-potential treatment of the $X^+$ state of He$_2^+$ by replacing, when calculating the reduced mass $\mu$,
         the nuclear masses $m(^N{\rm He}^{2+})$ ($N=3,4$) by effective atomic masses 
        \begin{equation}\label{red_masses}
        m_{\rm eff}(^N{\rm He}^{0.5+})= m(^N{\rm He}^{2+}) + 1.5 m_{\rm e}
        \end{equation}
        that equally distribute the masses of the three electrons among the two atoms. Similar approaches have been recommended, as discussed, {\it e.g.}, by LeRoy \cite{leroy17a}. Rather than using a potential-energy function calculated {\it ab initio}, we fit the parameters of an effective analytic potential-energy function $V_{\rm eff}(R)$ to best reproduce the available experimental data. 
        
        To accurately describe the level structure of the $X^+$ state of He$_2^+$ over the broad range of observed vibrational levels, a flexible function is needed that extends to large internuclear separations. We use the general potential form 
        \begin{equation}\label{eq:fullPotential}
                    V_{\rm eff}(R) = V_{\textrm{SR}}(R) (1 - \varPhi(R)) + V_{\textrm{LR}}(R) \varPhi(R),
        \end{equation}
        which gradually merges a short-range potential $V_{\textrm{SR}}(R)$ with a long-range potential $V_\textrm{LR}(R)$ using a switch function $\varPhi(R)$ operating at intermediate distances. 
        
        The rovibrational energies $E_{v^+,N^+}^{\text{calc}}$ and wavefunctions $\psi_{v^+,N^+}(R)$ are obtained by solving the radial Schrödinger equation (in atomic units)
        \begin{align}\label{eq:radialSchroedinger}
                    \left[-\frac{1}{2\mu}\frac{\text{d}^2}{\text{d}R^2} + V_{\rm eff}(R) + \frac{N^+(N^++1)}{2\mu R^2}\right] \psi_{v^+,N^+}(R) \notag\\
                    = E_{v^+,N^+}^{\text{calc}} \psi_{v^+,N^+}(R)
        \end{align} 
        using a Legendre–Gauss–Lobatto discrete-variable-representation (DVR) method \cite{manolopoulos88a,telnov99a,rescigno00a}. The adjustable parameters of the potential are optimized by minimizing the weighted sum of squared residuals
        \begin{equation}
                    \chi^2 = \sum_{v^+, N^+} \left(\frac{E_{v^+, N^+}^{\text{calc}} - E_{v^+, N^+}^{\text{exp}}}{\sigma_{v^+, N^+}^{\text{exp}}}\right)^2,
        \end{equation}
        where $\sigma_{v^+, N^+}^{\text{exp}}$ denote the experimental uncertainties, as given, {\it e.g.}, in Table~\ref{tab:energy_levels_Xp} for the rovibrational levels of the $X^+$ state of $^4$He$_2^+$ determined in the present work. 
        
        In a series of initial test calculations with several of the model potentials advocated by LeRoy \cite{leroy17a}, we observed that the bound levels with $v^+$ up to 19 can be well described with short-range potentials extending up to $R\approx 6.3~a_0$ and that the positions of the levels with $v^+=22,23$ observed in the microwave electronic spectrum primarily depend on the form of the long-range potential $V_{\textrm{LR}}(R)$, as already pointed out in Ref.~\cite{carrington95b}. These observations enabled us to optimize the short- and long-range parts of the potentials separately in the first phase of the analysis.
        
        To describe the short-range part of the potential $V_{\rm SR}(R)$ and the strongly bound levels, we used the Hannover polynomial expansion \cite{knoeckel04a,salumbides06a}
        \begin{equation}\label{eq:shortRange}
                    V_\textrm{SR}(R) = \sum_{i=0}^{n} a_i \cdot X^i(R), \quad \text{with } X(R) = \frac{R - R_\textrm{m}}{R + b \cdot R_\textrm{m}},
        \end{equation}
        in which the expansion coefficients $a_i$ and the parameter $b$ are adjustable parameters and $R_\textrm{m}$ is an arbitrary distance that needs to be chosen near the expected equilibrium separation $R_{\rm e}$. To fulfill this condition, we set $R_\textrm{m}$ to the equilibrium distance (2.042~$a_0$) reported by Tung et al. \cite{tung12a}. Ten expansion coefficients ($i=0-9)$ were sufficient to reproduce all observed levels of $^4$He$_2^+$ with $v^+\leq 19$ within their experimental uncertainties.
        
        The long-range potential $V_\textrm{LR}(R)$ corresponds to the electrostatic interaction series
        \begin{equation}\label{eq:longRange}
                    V_\textrm{LR}(R) = - \sum_{i \in m_i} \frac{C_i}{R^{i}}, \quad \text{with } m_i = [4, 6, 7, 8, 9,...],
        \end{equation}
        between He$^+$ and He. The coefficients of the first terms of this series, with $i$ up to 9, are accurately known from previous work \cite{davidson66, bishop89a, yan96a, yan98a, tung12a,gebala23a}.  For instance, the first term, in $R^{-4}$, represents the electrostatic interaction between the charge of the ion and the induced electric dipole moment of the neutral atom with $C_4 = \frac{\alpha_1}{2}$, where $\alpha_1$ corresponds to the static polarizability volume of the neutral He atom. The {\it ab initio} potential of Tung {\it et al.} \cite{tung12a} is indistinguishable from the potential obtained with Eq.~(\ref{eq:longRange}) for $i$ up to 9 at $R$ values beyond 17 $a_0$ and was shown to reproduce the experimental data on the $v^+=22$ and 23 levels with high accuracy. To extend $V_\textrm{LR}(R)$ to smaller $R$ values, higher terms of the long-range expansion series are needed. In order to retain an accurate description of the highest vibrational levels, we first adjusted the coefficients of the terms with $i=10$-19 to the {\it ab initio} potential of Tung {\it et al.} in a least-squares fit over the range of internuclear distances between 6.1 and 100~$a_0$, keeping the values of the coefficients $C_4-C_9$ fixed. In the following, we denote the fitted coefficients $K_i$ because there is no guarantee that the fitted coefficients exactly correspond to the "true" $C_i$ values.
        
        The short- and long-range parts of the potential were then smoothly joined using the switch function 
        \begin{equation}\label{eq:switch}
                    \varPhi(R) = \frac{1}{2}\left(1 + \tanh\left(\frac{R - R_\textrm{c}}{\varGamma}\right)\right)
        \end{equation}
        with $R_\textrm{c}$ set to $6.15~a_0$ and $\varGamma$ to $0.05~a_0$. 
        This choice of $R_\textrm{c}$ and $\varGamma$ preserved the accurate description of the experimental data obtained in the ranges $v^+=0-19$ and $v^+=22,23$. 
        In a final fit, the parameters $a_i$ and $b$ of $V_\textrm{SR}(R)$ were refined using the positions of the vibrational levels with $v^+=20$ and 21 located within 450~cm$^{-1}$ of the dissociation limit, as obtained using the {\it ab initio} potential of Tung {\it et al.} \cite{tung12a}.
        The optimal values of the fitted parameters and coefficients of the effective potential-energy function are listed in \Cref{tab:potential}. 
        The residuals (${\rm experimental} - {\rm calculated}$; $\Delta$) between the experimental rovibrational term values and the values calculated using the optimized  potential-energy function $V_{\rm eff}(R)$ are listed in Table~\ref{tab:energy_levels_Xp} and are all within the experimental uncertainties.
        
        The optimal effective potential-energy function $V_{\rm eff}(R)$ of the $X^+$ state of He$_2^+$ is depicted in Fig.~\ref{fig:FIG6_PEC_e-c_comparison}a, which also displays the corresponding vibrational energies and wavefunctions. 
        Fig.~\ref{fig:FIG6_PEC_e-c_comparison} also depicts, in panels b and c, the residuals between the experimental and calculated positions as determined by M\'atyus in calculations considering nonadiabatic, relativistic and quantum-electrodynamics corrections  before \cite{matyus18a} and after \cite{matyus25a} improving the theoretical treatment. 
        The agreement with the latter calculations is remarkable and similar to the agreement obtained with the effective potential energy function derived in this work, as shown in Fig.~\ref{fig:FIG6_PEC_e-c_comparison}d. 
        In particular, the residuals in Fig.~\ref{fig:FIG6_PEC_e-c_comparison}c and d reveal the same trends, which suggests that the deviations are dominated by the uncertainties of the measured term values. 
        The term values and binding energies of all 409 bound levels of the $X^+$ electronic ground state of $^4$He$_2^+$ predicted by the effective potential derived in the present work are provided in Tables~S2 and S3 of the supplemental material. 
        They may assist in experimental searches of the rovibrational levels of the $X^+$ state that could not be observed yet, in particular the rotational levels of the $v^+=20$ and 21 vibrational states. The calculations suggest that the $X^+(24,N^+=1)$ is bound. However, its binding energy is only $hc \times 0.0072\ {\rm cm}^{-1}$, which is less than the uncertainty in the dissociation energy $D_{\rm e}$ of the $X^+$ state (see Table~\ref{tab:potential}). Because the calculations of level energies with our effective potential relies on a simplified treatment of the reduced mass [see Equation~(\ref{red_masses})] and that only rotational levels with $N^+\leq 21$ were used when fitting the potential parameters, one cannot expect the positions of rotational levels with $N^+\gg 21$ to be predicted as reliably as is the case for $N^+\leq 21$. We estimate that the positions of these highest-$N^+$ levels could be in error by as much as 2 to 3~cm$^{-1}$.
        
        Table~\ref{tab:carrington} compares experimental term values and transition wavenumbers determined in the different isotopomers of He$_2^+$ with the corresponding values obtained from the effective potential-energy function derived in the present work.
        The upper, middle and bottom parts of the table concern term-value differences between the (22,5), (23,1) and (23,3) weakly bound levels of $^4$He$_2^+$ determined experimentally by Carrington {\it et al.} \cite{carrington95b}, infrared $R$-Type branch transition wavenumbers of the $v^+=1\leftarrow v^+=0$ vibrational band of $^4$He$^3$He$^+$ measured by Yu and Wing \cite{yu87a,yu88b}, and rotational term values in the ground vibrational level of $^3$He$_2^+$ reported by Raunhardt {\it et al.} \cite{raunhardt08a}, respectively. 
        Whenever available, the wavenumbers calculated by Tung {\it et al.} \cite{tung12a} are also listed for comparison. Whereas the experimental and calculated wavenumbers of the IR transitions of $^4$He$^3$He$^+$ agree within the experimental uncertainties in almost all cases, the calculated rotational term values in the ground state of $^3$He$_2^+$ are slightly larger (by up to $2\sigma$) than the experimental ones. 
        This discrepancy could either indicate a larger contribution from nonadiabatic effects in the lighter $^3$He$_2^+$ isotopomer than in the other isotopomers, or be caused by the correlation between the rotational constants of the $X^+$ state of $^3$He$_2^+$ and the $a$ state of $^3$He$_2$ in the fit of the lines observed in the PFI-ZEKE PE spectrum of $^3$He$_2$ in Ref.~\cite{raunhardt08a}. 
        Indeed, in the fit, the rotational constants of the $a$ state were kept fixed at values scaled from the corresponding constants of $^4$He$_2$. Nevertheless, the overall agreement is remarkable given that the effective potential-energy function was derived exclusively from data obtained for $^4$He$_2^+$.
        
        \begin{table}[ht!]
            {\centering
            \caption{Comparison of calculated and experimental term value differences (in cm$^{-1}$) between the three highest bound rovibrational states of the $X^+$ $^2\Sigma_u^+$ state of $^4$He$_2^+$ (top part of the table; experimental results from Ref.~\cite{carrington95b}), between the rotational levels of the lowest two vibrational levels of the $X^+$ $^2\Sigma_u^+$ state of $^3$He$^4$He$^+$ (middle part of the table; experimental results from Ref.~\cite{yu87a}), and between the first rovibrational levels of the $X^+$ $^2\Sigma_u^+$ state of $^3$He$_2^+$ (bottom part of the table; experimental results from Ref.~\cite{raunhardt08a}). 
            The numbers in parentheses indicate the estimated uncertainties, expressed in units of the last digit.
            See text for details.
            }
            \begin{ruledtabular}
            \begin{tabular}{l|ccc}
                & Expt. & Eff. Potential & Tung {\it et al.} \\ 
                &  & {\small This work} & {\small Ref. \cite{tung12a}} \\ \midrule
                $E_{(23,3)\leftarrow(23,1)}$ & 2.001\footnote{From Ref. \cite{carrington95b}} & 2.002 & 2.0020(10) \\
                $E_{(23,3)\leftarrow(22,5)}$ & 5.248$^\textrm{a}$ & 5.258 & 5.260(12) \\\midrule
                $E_{(1,2)\leftarrow(0,1)}$   &1781.8394(6)\footnote{From Ref. \cite{yu87a}}&1781.8398&1781.835\\
                $E_{(1,4)\leftarrow(0,3)}$   &1810.7172(6)$^\textrm{b}$&1810.7179&1810.713\\
                $E_{(1,5)\leftarrow(0,4)}$   &1824.2118(6)$^\textrm{b}$&1824.2126&1824.207\\
                $E_{(1,6)\leftarrow(0,5)}$   &1837.0549(6)$^\textrm{b}$&1837.0555&1837.051\\
                $E_{(1,7)\leftarrow(0,6)}$   &1849.2293(6)$^\textrm{b}$&1849.2300&1849.225\\
                $E_{(1,8)\leftarrow(0,7)}$   &1860.7190(6)$^\textrm{b}$&1860.7197&1860.715\\
                $E_{(1,10)\leftarrow(0,9)}$  &1881.5800(6)$^\textrm{b}$&1881.5798&1881.577\\
                $E_{(1,11)\leftarrow(0,10)}$ &1890.9155(6)$^\textrm{b}$&1890.9184&1890.917\\
                $E_{(1,12)\leftarrow(0,11)}$ &1899.5091(6)$^\textrm{b}$&1899.5086&1899.509\\\midrule
                $E_{(0, 1)\leftarrow(0,0)}$ &   18.773(18)\footnote{Calculated using molecular constants reported in Table II of Ref. \cite{raunhardt08a}}&18.7992  &18.798 \\
                $E_{(0, 2)\leftarrow(0,0)}$ &     56.30(5)$^\textrm{c}$&56.3756  &56.372 \\
                $E_{(0, 3)\leftarrow(0,0)}$ &   112.53(11)$^\textrm{c}$&112.6853 &112.677 \\
                $E_{(0, 4)\leftarrow(0,0)}$ &   187.41(18)$^\textrm{c}$&187.6624 &187.649 \\
                $E_{(0, 5)\leftarrow(0,0)}$ &    280.8(3)$^\textrm{c}$&281.2196  &281.199 \\
                $E_{(0, 6)\leftarrow(0,0)}$ &    392.7(4)$^\textrm{c}$&393.2477  &393.219 \\
                $E_{(0, 7)\leftarrow(0,0)}$ &    522.9(5)$^\textrm{c}$&523.6165  &523.578 \\
                $E_{(0, 8)\leftarrow(0,0)}$ &    671.3(7)$^\textrm{c}$&672.1748  &672.125 \\
                $E_{(0, 9)\leftarrow(0,0)}$ &    837.6(8)$^\textrm{c}$&838.7506  &838.689 \\
                $E_{(0, 10)\leftarrow(0,0)}$ &  1021.8(1.0)$^\textrm{c}$&1023.1516&1023.077 \\
                $E_{(0, 11)\leftarrow(0,0)}$ &  1223.5(1.2)$^\textrm{c}$&1225.1658&1225.076 \\
            \end{tabular}\\
            \label{tab:carrington}
            \end{ruledtabular}
            }

        \end{table}
        
    \subsection{Shape resonances of $^4$He$_2^+$}\label{shape_resonances}
        The representation of the rovibrational structure of the electronic ground state of He$_2^+$ through an effective potential-energy function enables the straightforward calculation of the positions and widths of the shape resonances. 
        Only a few of these resonances, corresponding to large values of the rotational-angular-momentum quantum number $N^+$ and relatively low values of the vibrational quantum number $v^+$, have been observed in an ion beam experiment by monitoring the momentum of the He$^+$ fragments produced by tunneling through the centrifugal barrier \cite{maas76a}. 
        Their assignment was only possible by comparisons with predictions from {\it ab initio} calculations \cite{maas76a,bauschlicher89a,cencek95a}, considering that only shape resonances with lifetimes around 1~$\mu$s could be detected experimentally. 
        Whereas the different calculations mostly agree on the assignment of the resonances observed experimentally, the calculated positions differ by up to 100~cm$^{-1}$, primarily because of uncertainties in the dissociation energy of He$_2^+$. 
        In the case of $^4$He$_2^+$ four shape resonances were observed, with central positions of $23\pm 1$, $406\pm 7$, $968\pm 13$ and $2882 \pm 44$~cm$^{-1}$ \cite{maas76a}, and were assigned to states with ($v^+,N^+)$ quantum numbers of (19,17), (12,37) or (13,35), (8,47) or (9,45), and (1,63), respectively \cite{maas76a,bauschlicher89a,cencek95a}. 
        
        \begin{table}[ht]
            \centering
            \caption{Predicted positions and widths (both given in cm$^{-1}$) of shape resonances of $^4$He$_2^+$ using the empirical effective potential-energy function determined in this work for $N^+\leq 29$.}
            \begin{ruledtabular}
            \begin{tabular}{l@{\hspace{-.5pt}}cc|l@{\hspace{-.5pt}}cc}
                $v^+,N^+$ & {$E_\textrm{res}$} & {$\varGamma$} &
                $v^+,N^+$ & {$E_\textrm{res}$} & {$\varGamma$} \\
                \midrule
                15,29 & 168.79 & 2.26$\times 10 ^{-9}$ & 18,23 & 198.82 & 9.64 \\
                16,25 & 0.88 & 9.17$\times 10 ^{-3 }$  & 19,17 & 30.48 & 8.79$\times 10 ^{-7 }$ \\
                16,27 & 195.35 & 1.47$\times 10 ^{-4 }$ & 19,19 & 108.98 & 3.28 \\
                16,29 & 363.30 & 5.78 & 20,15 & 52.39 & 1.60 \\
                17,23 & 59.12 & 3.84$\times 10 ^{-11}$ & 21,11 & 20.40 & 1.42 \\
                17,25 & 207.71 & 2.42$\times 10 ^{-1 }$ & 22,7 & 4.34 & 6.27$\times 10 ^{-1 }$ \\
                18,21 & 97.12 & 2.01$\times 10 ^{-3 }$ & & & \\
            \end{tabular}
            \label{tab:Low_N_resonances}
            \end{ruledtabular}

        \end{table}
        The effective potential-energy function derived in the present work, with an estimated uncertainty of the dissociation energy of only 0.1~cm$^{-1}$ provided an opportunity to recalculate the shape resonances of He$_2^+$, as done previously for the shape and orbiting resonances of H$_2^+$ \cite{beyer16a,beyer17a,beyer18b}.
        Using an algorithm based on the log-derivative formalism \cite{landau08a,johnson77a}, as detailed in the Appendix, we determined the positions of all shape resonances with widths larger than 10$^{-12}$~cm$^{-1}$. 
        In some cases, the resonance widths could not be determined reliably because of their extreme sharpness. A total of 74 resonances were identified and their positions and widths are listed in Table S4 of the supplementary material. 71 of these resonances have their central energy below the maximum of the centrifugal barrier and three of them just above and represent orbiting resonances. Because most of these resonances are associated with high values of $N^+$, their positions may not be predicted as reliably as the positions of low-$N^+$ bound states, as explained above.
        \begin{table*}[ht!]
            \centering
            \caption{Positions and widths of the shape resonances of $^4$He$_2^+$ predicted using the empirical effective potential-energy function determined in this work and selected to be closest in energy to resonances experimentally observed by Maas {\it et al.} \cite{maas76a} and compared to the theoretical values of Cencek {\it et al.} \cite{cencek95a}, Bauschlicher {\it et al.} \cite{bauschlicher89a} and Maas {\it et al.} \cite{maas76a}.
            The positions $E_\textrm{res}$ are given in cm$^{-1}$ with respect to the dissociation limit.
            The widths $\varGamma$, also given in cm$^{-1}$, of the resonances are taken as the FWHM of the peak observed when expressing the derivative of the phase shift as function of the resonance energy (see appendix for details).
            }\label{resonances_comparison}
            \begin{ruledtabular}
                
            \begin{tabular}{cccccccccc}
                & {Expt.\footnote{Experimental and theoretical values from Maas {\it et al.} \cite{maas76a}.}} & \multicolumn{2}{l}{Eff. Potential\footnote{Effective potential values from this work.}}& \multicolumn{2}{l}{Cencek {\it et al.}\footnote{Theoretical values from Bauschlicher {\it et al.} \cite{bauschlicher89a}.}} &   \multicolumn{2}{l}{Bauschlicher {\it et al.}\footnote{Theoretical values from Cencek and Rychlewski {\it et al.} \cite{cencek95a}.}} & \multicolumn{2}{l}{Maas {\it et al.}$^\textrm{a}$} \\
                {$v^+,N^+$} & {$E_\textrm{res}$} & {$E_\textrm{res}$} & {$\varGamma$} & {$E_\textrm{res}$} & {$\varGamma$} & {$E_\textrm{res}$} & {$\varGamma$} & {$E_\textrm{res}$} & {$\varGamma$}\\
                \midrule
                19,17 & 23(1) & 30   & 9$\times 10 ^{-7}$ & 32 & 1.8$\times 10 ^{-6}$  &   &     & & \\
                12,37 & 406(7) & 411  & 8$\times 10 ^{-9}$  &   & & 462  & 2.3$\times 10 ^{-7}$ & 362 & 1.3$\times 10 ^{-10}$ \\
                13,35 &  & 406  & 4$\times 10 ^{-6}$ & 409 & 5.2$\times 10 ^{-6}$ & 456  & 1.0$\times 10 ^{-4}$ & & \\
                8,47  & 968(13) & 1012 & 2.5$\times 10 ^{-7}$ & 1017 & 2.9$\times 10 ^{-7}$ & 1065 & 1.1$\times 10 ^{-6}$ & 981 & 1.4$\times 10 ^{-7}$ \\
                9,45  &  & 933  & 5$\times 10 ^{-6}$ & 937 & 6.3$\times 10 ^{-6}$ & 985  & 2.2$\times 10 ^{-5}$ & & \\
                1,63  & 2882(44) & 2895 & 1.2$\times 10 ^{-6}$ & 2900 & 1.2$\times 10 ^{-6}$ & 2936 & 1.1$\times 10 ^{-6}$ & 2895 & 3.3$\times 10 ^{-6}$ \\
            \end{tabular}
                \label{tab:Experimental_resonances}
            \end{ruledtabular}

        \end{table*}
        Table~\ref{tab:Low_N_resonances} lists the positions of all shape resonances with $N^+$ values below $31$, corresponding to the range of rotational states that can be accessed from He$_2^\ast$ generated with an dielectric-barrier-discharge source. 
        Only one of these, the (19,17) resonance observed at $23\pm 1$~cm$^{-1}$, was previously tentatively assigned by Cencek and Rychlewski, who predicted a resonance position of 32~cm$^{-1}$ and a resonance width of $1.8\times 10^{-6}$~cm$^{-1}$ (see Table~\ref{tab:Experimental_resonances}). 
        Our calculations predict a position of 30~cm$^{-1}$ and a width of $0.9\times 10^{-6}$~cm$^{-1}$ for this resonance.

        Table~\ref{tab:Experimental_resonances} presents the positions and widths of the shape resonances that are closest in positions and widths to those observed experimentally by Maas {\it et al.} \cite{maas76a}. 
        This selection corresponds closely to the assignments proposed in earlier theoretical studies \cite{maas76a,bauschlicher89a,cencek95a}. 
        Our values are closest to the values reported by Cencek and Rychlewski \cite{cencek95a}. 
        Whereas the assignments of the resonances observed at 406(7) and 2882(44)~cm$^{-1}$ to the (13,35) and (1,63) resonances, respectively, seem plausible, deviations by more than $3\sigma$ in the case of the resonances observed at 23(1) and 968(13)~cm$^{-1}$ remain unsatisfactory. 
        New measurements of the shape resonances of He$_2^+$ would be desirable.

\section{Conclusions}

In this article, we have demonstrated an experimental approach, 
based on a two-photon excitation scheme via selected rotational levels of the $c(v^\prime=3,4)$ levels that are subject to predissociation by quantum-mechanical tunneling, to access a broad range of vibrational levels of the $X^+$ electronic ground state of He$_2^+$. A large number of rovibrational levels of this state, with vibrational quantum numbers $v^+$ up to 19, were characterized by high-resolution PFI-ZEKE PE spectroscopy, which also provided detailed information on the photoionization dynamics. By combining these new results with results obtained in previous work on the lowest ($v^+=0-2$) and highest ($v^+=22,23$) vibrational levels of the $X^+$ state, we determined an effective potential-energy function that accurately describes the observed level structure not only of $^4$He$_2^+$, but also of $^4$He$^3$He$^+$ and $^3$He$_2^+$. The dissociation energy of He$_2$ was determined to be $D_{\rm e}=19\,956.10(10)$~cm$^{-1}$ [$D_0(^4{\rm He}_2^+)=19\,101.29(10)$~cm$^{-1}$].

This effective potential-energy function was used to calculate the positions of all bound levels of the electronic ground state of He$_2^+$. A total of 409 rovibrational levels with $v^+$ values up to 24 and $N^+$ values up to 67 was identified. With the same potential, 74 shape resonances associated with the  $X^+$ state of $^4$He$_2^+$ were predicted and used in a discussion of the assignment of the four shape resonances observed experimentally \cite{maas76a}.

\section{Acknowledgement}
We thank Josef A. Agner and Hansj{\"u}rg Schmutz (both ETH Zurich) for technical support and Professor Edit M\'atyus (E{\"o}tv{\"o}s Lor{\'a}nd University, Budapest) for very useful discussions and for providing the results of her latest calculations prior to publication.
This work is supported by the Swiss National Science Foundation under project 200021-
236716.

\appendix*

\section{Calculation of the shape resonances}
        
	To calculate the shape resonances (quasi-bound states) above the dissociation limit of the $X^+$ electronic state of $^4$He$_2^+$, we used an algorithm based on the log-derivative formalism \cite[Ch. 4.2]{landau08a}\cite{johnson77a}. Because the potential-energy function $V_{\rm eff}(R)$ goes to zero faster than $R^{-2}$, the solutions $\chi_{k,N^+}(R)$ of the time-independent Schr{\"o}dinger equation [Eq.~(\ref{eq:radialSchroedinger})] have to obey the asymptotic condition
	\begin{equation}\label{schr_asym}
		\left\{R^2\dv[2]{R} - N^+(N^++1) + k^2R^2\right\} \chi_{k,N^+}(R)= 0,
	\end{equation}
	with 
	\begin{equation}
		k^2 = 2\mu\left[E-V_{\rm eff}(R=\infty)\right] > 0.
	\end{equation}
	The solutions of Eq.~(\ref{schr_asym}) are the Ricatti-Bessel functions $\hat{\jmath}_{N^+}(kR)$ and $\hat{y}_{N^+}(kR)$ of order $N^+$, where $N^+$ is the rotational-angular-momentum quantum number. In the asymptotic region, i.e., outside the range of $V_{\rm eff}$, $\chi_{k,N^+}(R)$ is given by
	\begin{equation}
		\chi_{k,N^+}(R) = A_{N^+} kR \bigg[\cos{\delta_{N^+}}\hat{\jmath}_{N^+}(kR) - \sin{\delta_{N^+}}\hat{y}_{N^+}(kR)\bigg],
	\end{equation}
	where $\delta_{N^+}$ is the scattering phase shift, and $A_{N^+}$ a normalization constant.
	
	Using the logarithmic-derivative expression
	\begin{equation}
		\mathcal{Y}(R) \equiv \frac{1}{\chi_{k,N^+}(R)} \dv{\chi_{k,N^+}(R)}{R} = \dv{\ln\{\chi_{k,N^+}(R)\}}{R},
	\end{equation}
	one obtains
	\begin{equation}\label{eq:PhaseShiftLOGD}
		\tan \delta_{N^+} = \frac{\hat{\jmath}_{N^+}^{\prime}(kR)-\hat{\jmath}_{N^+}(kR) \mathcal{Y}(R)}{\hat{y}_{N^+}^{\prime}(kR)-\hat{y}_{N^+}(kR) \mathcal{Y}(R)}.
	\end{equation}
		where the prime designates the derivative with respect to $R$, i.e.,
	\begin{equation}
		\hat{y}_{N^+}^{\prime}(kR) = \dv{R}\hat{y}_{N^+}(kR) = k\dv{(kR)}\hat{y}_{N^+}(kR).
	\end{equation}
	The scattering phase shift can be separated into a component $\delta_{N^+}^\mathrm{bg}(E)$ that slowly varies with the energy $E$ ("bg" for background) and a resonant contribution $\delta_{N^+}^\mathrm{res}(E)$ \cite{burke11a}
	\begin{equation}
		\delta_{N^+}(E) = \delta_{N^+}^\mathrm{bg}(E) + \delta_{N^+}^\mathrm{res}(E).
	\end{equation}
	Shape resonances correspond to $\delta_{N^+}^\mathrm{res}(E_\mathrm{res}) = \pi(n+1/2)$, where $n$ is an integer. Expanding around $E_\mathrm{res}$ (see Chapter 13.5 of Ref.~\cite{townsend00b}) leads to the expression
	\begin{align}
		\cot{\delta_{N^+}^\mathrm{res}(E)} &= \cot{\delta_{N^+}^\mathrm{res}(E_\mathrm{res})} - \frac{2}{\varGamma}(E-E_\mathrm{res}) + \mathcal{O}\left[(E - E_\mathrm{res})^2\right] \nonumber \\ 
		&= -\frac{2}{\varGamma}(E-E_\mathrm{res}) + \mathcal{O}\left[(E - E_\mathrm{res})^2\right] \label{eq:TaylorExpansionPhaseShift},
	\end{align}
	with 
	\begin{equation}
			\frac{2}{\varGamma} = -\left.\pdv{\cot{\delta_{N^+}^\mathrm{res}(E)}}{E}\right|_{E=E_\mathrm{res}}.
	\end{equation}
	\Cref{eq:TaylorExpansionPhaseShift} implies that, in the vicinity of a resonance, 
	\begin{equation}
		\delta_{N^+} = \delta_{N^+}^\mathrm{bg} + \arctan\left(\frac{\varGamma/2}{E_\mathrm{res}-E}\right).
	\end{equation}
	and therefore
	\begin{equation}\label{eq:PhaseShiftDerivative}
		\dv{\delta_{N^+}}{E} = \frac{\varGamma/2}{(E_\mathrm{res}-E)^2 + \varGamma^2/4}.
	\end{equation}
	To determine the positions and widths of the resonances, the energy is varied in discrete steps (index $i$) and, at each step $E_i$, the logarithmic derivative is integrated outwards \cite{johnson77a} and the phase shift is calculated using \Cref{eq:PhaseShiftLOGD}. 
	The resonance positions $E_\mathrm{res}$ correspond to the energy at which maxima of \Cref{eq:PhaseShiftDerivative} occur and their widths $\varGamma$ are determined by inserting $\delta_{N^+}(E_\mathrm{res})$ into \Cref{eq:PhaseShiftDerivative}, i.e.,
	\begin{equation}
		\dv{\delta_{N^+}(E_\mathrm{res})}{E} = \frac{2}{\varGamma}.
	\end{equation}

\bibliography{grbib} 

\end{document}